\providecommand{\printstyle}{reprint}
\providecommand{\floatlocation}{}

\documentclass[10pt,
	aip,apl,longbibliography,  
	\printstyle,
	\floatlocation,
	citeautoscript,
	superscriptaddress,
	nofootinbib,
	nobalancelastpage,
	byrevtex,
	floatfix]{revtex4-1}

\usepackage{import}
	
\usepackage[pdftex]{graphicx} 

\usepackage{times}
\usepackage[T1]{fontenc}

\usepackage{amsmath}
\usepackage{bm}
\usepackage{physics}
\usepackage{siunitx}
\usepackage[colorlinks=true,
    citecolor=blue,
    linkcolor=blue,
    urlcolor=blue,
    pagebackref=false]{hyperref}
    
\usepackage{array}
\newcolumntype{x}[1]{>{\centering\let\newline\\\arraybackslash\hspace{0pt}}m{#1}}

\newlength{\widthofonecolumn}
\newlength{\widthoftwocolumns}
\usepackage{fancyhdr}
\begin{document}

\title{Reverse Monte Carlo reconstruction of electron spin-label coordinates from scanned-probe magnetic resonance  microscope signals}

\author{Hoang Long Nguyen}
\affiliation{
	Department of Chemistry and Chemical Biology,
	Ithaca, New York 14853-1301}
\author{John A. Marohn}
\email{jam99@cornell.edu}
\affiliation{
	Department of Chemistry and Chemical Biology,
	Ithaca, New York 14853-1301}

\begin{abstract}
\label{abstract}
Individual electron spins have been observed using magnetic resonance in combination with a number of distinct detection approaches.
The coordinates of an individual electron spin can then in principle be determined by introducing a \SI{10}{} to \SI{100}{\nano\meter} diameter magnetic needle, scanning the needle, and collecting signal as a function of the needle's position.
Although individual electrons have recently been localized with nanometer precision in this way using a nitrogen-vacancy center in diamond as the spin detector, the experiment's low signal-to-noise ratio limited acquisition to two-dimensional scanning, enabled the observation of just a few dozen data points, and was incompatible with nitroxide spin labels widely used to label proteins and nucleic acids.
To remedy these limitations, we introduce and numerically simulate a protocol for detecting and imaging individual nitroxide electron spins with high spatial resolution.
In our protocol, electron-spin magnetic resonance is detected mechanically: a scanned magnet-tipped cantilever is brought near the sample, modulated microwaves are applied to resonantly excite electron spins, and changes in spin magnetization are detected as a shift in the mechanical frequency of the cantilever.
By carefully applying resonant microwaves in short bursts in synchrony with the cantilever's oscillation, we propose to retain high spatial resolution even at large cantilever amplitude where sensitivity is highest.
Numerical simulations reveal nanometer-diameter rings of frequency-shift signal as the tip is scanned laterally and individual electrons come in and out of resonance.
Our primary finding is that it is possible --- using a Bayesian, reverse Monte Carlo algorithm introduced here --- to obtain the full three-dimensional distribution of electron coordinates from the signal rings revealed in a two-dimensional frequency-shift map. 
This reduction in dimensionality brings within reach, on a practical timescale, the angstrom-resolution three-dimensional imaging of spin-labeled macromolecules.
\end{abstract}
\date{\today}
\maketitle
\thispagestyle{fancy}

\section{Introduction}
\label{sec:Introduction}

Detecting and imaging the spin state of electrons in individual molecules or defects has been a long-sought goal of the chemical physics community \cite{Wrachtrup2016aug,Blank2017jul}.
Efforts to detect single electron spins  in specialized materials have been made using light \cite{Kohler1993may,Wrachtrup1993may,Koehler1994sep,Brouwer1996aug,Gruber1997jun,Grinolds2013feb, Grinolds2014mar,Shi2015mar,Sushkov2014nov} and current \cite{Manassen1989may,Durkan2002jan,Balatsky2012apr,Meier2008apr,Loth2010sep,Baumann2015oct, Xiao2004jul,Morello2010sep,Thiele2014jun} and, more generally, using quantum interference devices \cite{Levenson-Falk2013may,Granata2016feb,Levenson-Falk2016nov,Vasyukov2013sep}, inductive detection \cite{Artzi2015feb,Bienfait2015dec,Bienfait2017oct,Probst2017nov}, forces \cite{Sidles1991jun,Sidles1992feb,Rugar1992dec,Rugar2004jul}, and force gradients \cite{Garner2004jun,Moore2009dec}. 
Here we introduce a force-gradient protocol for imaging individual nitroxide spin labels, numerically simulate signal expected from a doubly labeled protein, and develop a reverse Monte Carlo algorithm that recovers individual electron-spin coordinates. 
The protocol employs force-gradient detection because of its generality, sensitivity, depth of view, and compatibility with the spin relaxation times of nitroxide spin labels.

Nearly twenty five years ago researchers first succeeded in observing the electron-spin state of an individual $\pi$-conjugated molecule in a frozen glass using optically detected magnetic resonance \cite{Kohler1993may,Wrachtrup1993may}; the nuclear-spin state of individual $^{13}\text{C}$ nuclei in the molecule were even detectable through their hyperfine coupling to the electron spin \cite{Koehler1994sep,Brouwer1996aug}.
The nitrogen vacancy in diamond has proven a fertile ground for optically polarizing and detecting the spin state of single electronic defects at room temperature \cite{Gruber1997jun}.
The nitrogen-vacancy center in diamond has been harnessed as a magnetometer for detecting and imaging nearby single electron spins \cite{Grinolds2013feb,Grinolds2014mar,Shi2015mar} and nuclear spins \cite{Sushkov2014nov}.
In parallel with these developments, numerous non-optical methods for detecting single spins have been explored.
Individual paramagnetic electron spins at a surface have been observed via scanning tunneling microscopy using both unmagnetized tips \cite{Manassen1989may,Durkan2002jan,Balatsky2012apr} and spin-polarized tips \cite{Meier2008apr,Loth2010sep,Baumann2015oct}.
The spin state of individual electrons at dopants in silicon \cite{Xiao2004jul,Morello2010sep} and individual nuclear spins in a molecular magnet \cite{Thiele2014jun} have been detected electrically. 
Superconducting quantum interference devices, if suitably miniaturized \cite{Levenson-Falk2013may,Granata2016feb,Levenson-Falk2016nov,Vasyukov2013sep} and carefully fabricated \cite{Vasyukov2013sep}, can achieve single electron-spin sensitivity \cite{Vasyukov2013sep}. 
Inductively detected electron spin resonance, harnessing advances in micrometer-sized resonators \cite{Artzi2015feb} and cryogenic amplifiers \cite{Artzi2015feb,Bienfait2015dec,Bienfait2017oct,Probst2017nov}, has reached a sensitivity of a few tens of electron spins. 
Magnetic resonance force microscopy (MRFM)\cite{Sidles1991jun,Sidles1992feb,Rugar1992dec,Rugar1994jun} was used by Rugar and coworkers to detect and image the electron-spin state of an individual dangling bond in quartz \cite{Rugar2004jul} and has enabled a number of groups to observe proton nuclear magnetization with a sensitivity of a few hundred spins and a spatial resolution of a few nanometers \cite{Degen2009feb,Longenecker2012nov,Nichol2013sep,Moores2015may,Rose2017jul}.

Nanometer-scale imaging of electron spin density was achieved in MRFM \cite{Rugar2004jul} and nitrogen-vacancy experiments \cite{Grinolds2013feb,Grinolds2014mar} by  employing a scanned submicrometer-diameter magnetic pillar or needle to make the magnetic resonance condition dependent on position.
The spatial resolution in such a magnetic resonance imaging experiment is determined not by the radius of the scanned magnetic tip but by the tip's field gradient and the sample's magnetic resonance linewidth.
In the Ref.~\citenum{Grinolds2014mar} experiment, for example, the linewidth was $\Delta B = \SI{0.3}{\milli\tesla}$ and the gradient was $G = \SI{1.2}{\milli\tesla \per \nano\meter}$. 
The calculated linewidth-limited magnetic resonance imaging resolution was $\Delta B/G = \SI{0.25}{\nano\meter}$; the observed resolution,  $0.8$ to $\SI{1.5}{\nano\meter}$, was limited by the available signal-to-noise ratio.
A number of particularly intense magnetic field gradient sources have been developed for MRFM experiments: off-tip magnetic pillars made of FeCo \cite{Degen2009feb} and Dy \cite{Mamin2012jan}, a switchable current flowing through a \SI{100}{\nano\meter} constriction \cite{Nichol2012feb,Nichol2013sep,Rose2017jul}, and a switchable commercial read-write head \cite{Tao2016sep}.
On-tip cobalt nanorods \cite{Longenecker2011may,Longenecker2012nov} have been developed that produce a magnetic field gradient as large as \SI{5}{\milli\tesla\per\nano\meter}.

Moore, Marohn, and coworkers introduced the idea of determining the tertiary structure of a single copy of a frozen biomolecule or biomolecular complex by affixing nitroxide labels to it and mapping the locations of the individual nitroxide electron spins using MRFM \cite{Moore2009dec}.
Detecting a single nitroxide by MRFM is challenging, however.
To achieve single-spin sensitivity the experiment of Rugar \emph{et al}.\ \cite{Rugar2004jul} required spin-locking the sample magnetization for nearly a second, clearly incompatible with the nitroxide's short relaxation times ($T_{1\mathrm{e}} = \SI{1}{\milli\second}$ and $T_{2\mathrm{e}} = \SI{450}{\nano\second}$ at \SI{4.2}{\kelvin}).
Moore and coworkers showed that electron-spin resonance could be detected in a force-gradient MRFM experiment in which a periodic spin flip in the sample modulated the mechanical oscillation frequency of a nearby magnet-tipped cantilever \cite{Garner2004jun}.
This approach opens up a new route for pushing MRFM to single-electron sensitivity because, in contrast with the Ref.~\citenum{Rugar2004jul} experiment, a force-gradient experiment is not limited to observing spin-force fluctuations (which have random sign) but instead detects the sample's average Curie-law magnetization (which has a well-defined sign).

Since Moore's experiment \cite{Moore2009dec}, MRFM's per-spin sensitivity and achievable electron-spin polarization have improved significantly.
The Moore \emph{et al}.\ experiment was carried out at a temperature of $\SI{4.2}{\kelvin}$, in a magnetic field of $\SI{0.6}{\tesla}$, and employed a high-compliance cantilever having a sensitivity of $\SI{7.5}{\atto\newton}/\sqrt{\si{\hertz}}$.
The experiment's cantilever had a radius $r_{\mathrm{tip}} = \SI{2}{\micro\meter}$ nickel tip attached and read out the cantilever frequency using a tip oscillation amplitude of $x_{\text{p}-\text{p}} = \SI{330}{\nano\meter}$.
The experiment achieved a sensitivity of $400 \: \text{spins}/\sqrt{\si{\hertz}}$.
In Ref.~\citenum{Longenecker2012nov}, Longenecker and coworkers prepared a cantilever similar to Moore's with a cobalt nanorod tip that was $\SI{225}{\nano\meter}$ wide and $\SI{79}{\nano\meter}$ thick; the lateral gradient at a distance of $13$ \si{\nano\meter} below the bottom of the tip was estimated to be $5.5$ to $8.3$ \si{\milli\tesla/\nano\meter}. 
The sensitivity reported for this cantilever was $500$ proton magnetic moments in a 1 \si{\milli\hertz} bandwidth. 
As the electron spin's magnetic moment is 660 times larger than that of a proton, this $500$-proton sensitivity is essentially equivalent to single electron sensitivity, motivating us to consider an individual electron imaging experiment. 
Carrying out the Moore experiment with the nanorod-tipped cantilever of Ref.~\citenum{Longenecker2012nov} would, we predict, yield single-electron-spin sensitivity in just a few seconds of averaging.
However, the large cantilever amplitude required to precisely read out the single-spin cantilever frequency shift would blur the image, severely limiting the achievable spatial resolution.

The present study has two goals.
Below we introduce an enhanced spin-detection protocol that retains the advantages of Moore \emph{et al}.'s approach while enabling nanometer-resolution imaging of single-electron magnetization in a scanned-tip experiment.
Maps of spin frequency-shift signal versus tip position are numerically calculated for a doubly spin-labeled individual protein observed with the new protocol and, for comparison, the Moore protocol.   
We then introduce a rigorous and facile Bayesian, reverse Monte Carlo algorithm for reconstructing the distribution of spin-label coordinates from the simulated spin frequency-shift maps.
We finally introduce an approximate non-iterative reconstruction algorithm,  employing Fourier deconvolution stabilized by Tikhonov regularization, which provides a starting point for the reverse Monte Carlo algorithm. 
The new approach dramatically reduces the number of free parameters, significantly improves the image-reconstruction convergence time, and naturally outputs the desired three-dimensional distribution of electron coordinates.

The paper is organized as follows. 
In Sec.~\ref{sec:detection} we present a new protocol for imaging individual electron spins using a magnet-tipped cantilever, force-gradient detection, and intermittently applied microwaves.
In Sec.~\ref{sec:simulation} we derive a Bloch-like solution for the steady-state electron-spin magnetization in such an intermittent-irradiation experiment.
We then present the numerical simulation of a two-dimensional (2D) scanned force-gradient signal from a doubly labeled biomolecule acquired using the Sec.~\ref{sec:detection} protocol and assuming modest improvements in existing MRFM technology. 
In Section \ref{sec:Bayesian} we present a reverse Monte Carlo reconstruction scheme and use it to obtain the three-dimensional (3D) coordinates of the sample's two electron spins from the simulated 2D cantilever frequency-shift map. 
Finally, in Section \ref{sec:Tikh}, we introduce a second reconstruction scheme using Fourier deconvolution with Tikhonov regularization that obtains an approximate 3D spin density map non-iteratively. 
The spins' coordinates extracted from this spin density map then serve as the starting point for the reverse Monte Carlo reconstruction, reducing the reconstruction time by an order of magnitude.

\section{Magnetic Resonance Mechanical Detection protocol}
\label{sec:detection}

\begin{figure}
\includegraphics
	[width=0.80\widthofonecolumn]
	{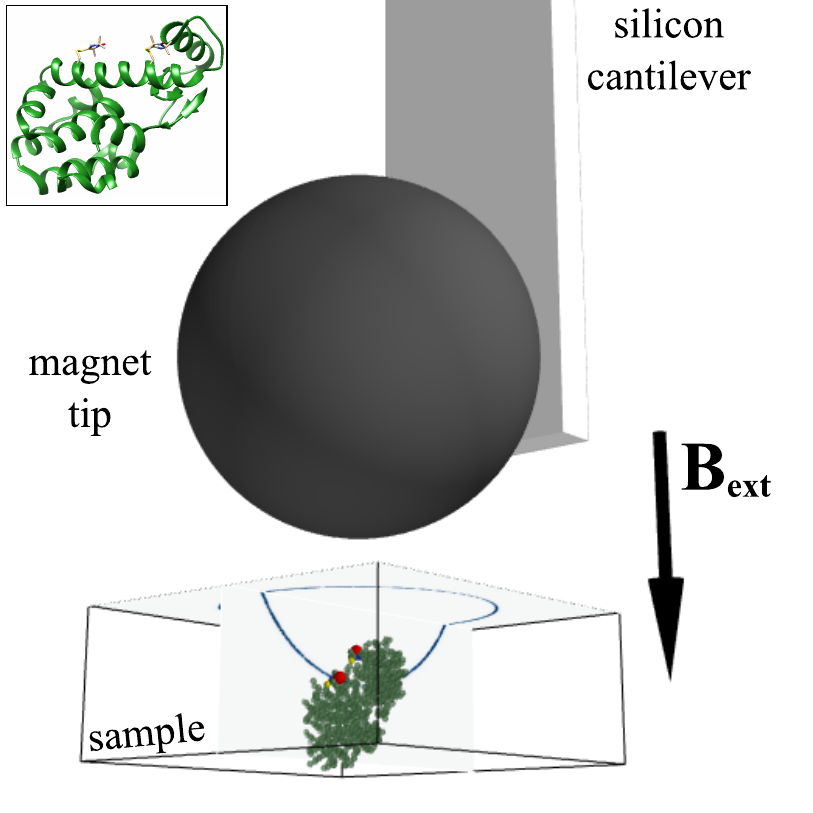}
\caption{Electron-spin resonance experiment schematic showing the relative orientation of the applied magnetic field ${\bm{B}}_{\mathrm{ext}} = B_{\mathrm{ext}} \: \hat{z}$, 
	cantilever oscillating in the $\hat{x}$ direction (gray rectangle),
	magnetic tip (black sphere), and 
	T4 lysozyme molecule (green ribbon diagram) with two nitroxide spin probes attached (red spheres).
	A microwave source (not shown) excites a resonant slice of electron spins in the molecule (blue cutaway region).
	The lysozyme molecule is magnified by $10\times$ for clarity; see the expanded ribbon diagram in the upper-left inset.}
\label{fig:3K2R-PSF}
\end{figure}

Here we numerically simulate the force-gradient magnetic resonance experiment sketched in Fig.~\ref{fig:3K2R-PSF}.
The sample consists of a single mutant T4 lysozyme protein with two nitroxide electron-spin labels affixed to it (pdb 3K2R, spin-labeled T4 lysozyme mutant  K65V1/R76V1);
the spin labels are $20 \: \text{\AA}$ apart.
An ultrasensitive silicon cantilever is prepared with a radius $r_{\mathrm{tip}} = \SI{75}{\nano\meter}$ spherical cobalt tip.
The cantilever is brought to a tip-sample separation of $h = \SI{60}{\nano\meter}$.
A magnetic field is applied parallel to the long axis of the cantilever to polarize the ferromagnetic tip and the electron spins in the sample below.
A transverse oscillating microwave magnetic field (not shown) is applied to saturate or invert the sample's electron spins. 
Because of the tip's large magnetic field gradient, electron spins meet the magnetic resonance condition only in a thin, hemispherical constant-field region of space below the tip; we term this region the \emph{resonant slice} (Fig.~\ref{fig:3K2R-PSF}, blue cutaway region).

\begin{figure}
\includegraphics
	[width=\widthofonecolumn]
	{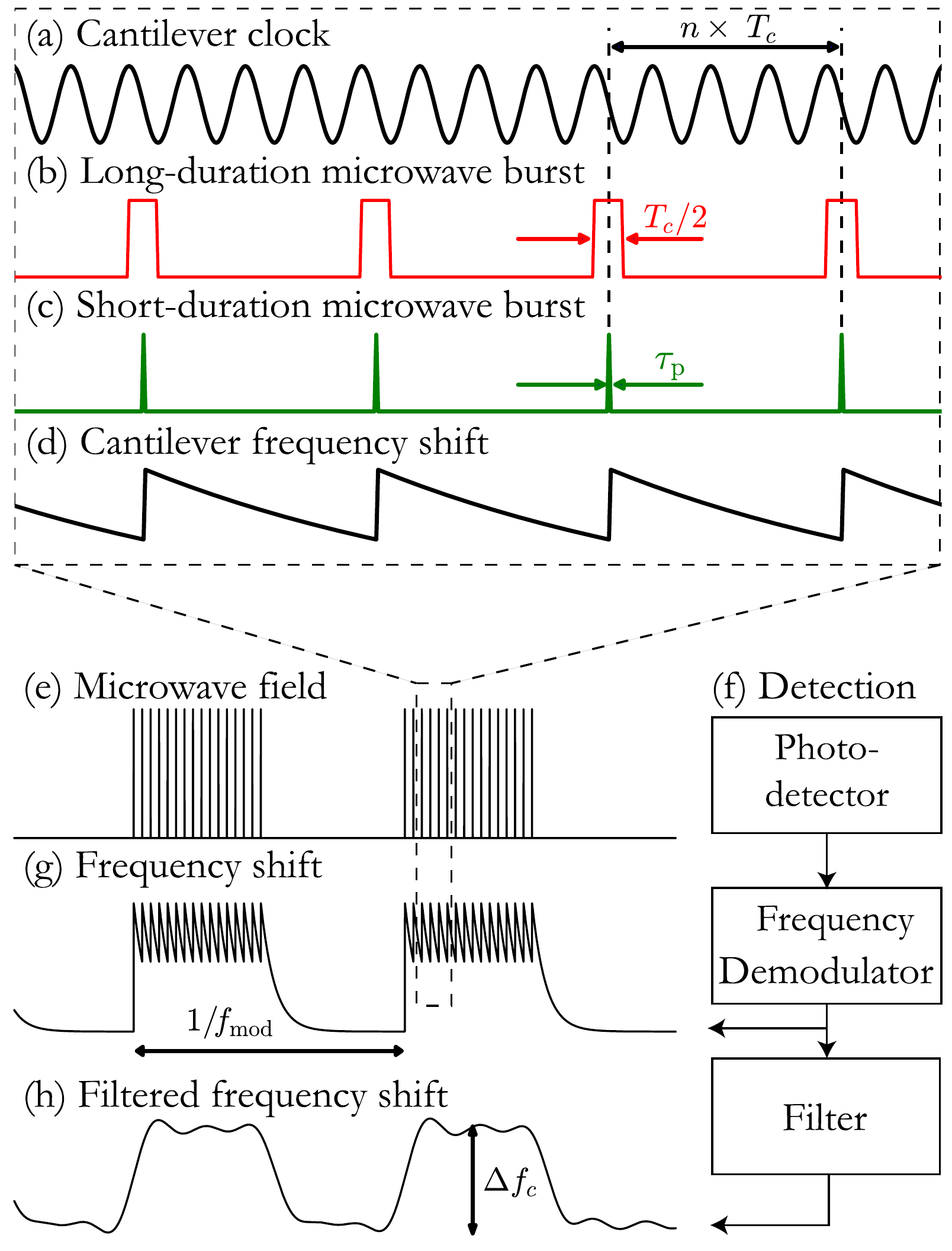}
\caption{Protocol for mechanically detecting and imaging the spin magnetization of a single electron. 
	(a) The Fig.~\ref{fig:3K2R-PSF} cantilever is oscillated at its mechanical resonance frequency ($f_c = \SI{6.5}{\kilo\hertz}$; cantilever period $T_c = \SI{154}{\micro\second}$).
A saturating microwave burst is applied every $n = 4$ cantilever cycles in synchrony with cantilever zero crossings: 
	(b) a long-duration burst (red, $\tau_{\mathrm{p}} = T_c/2 = \SI{77}{\micro\second}$, adapted from Ref.~\citenum{Moore2009dec}) and 
	(c) a short-duration burst (green, $\tau_{\mathrm{p}} = \SI{0.8}{\micro\second}$).
	(d) A microwave-induced change in electron spin magnetization modifies the cantilever's spring constant and shifts its resonance frequency.
	(e) The microwave bursts are modulated in an on-off pattern with the \emph{on} and \emph{off} periods each lasting $t_{\mathrm{mod}} = \SI{9.8}{\milli\second}$.
	(f) The cantilever oscillation is observed with a photodetector whose output is passed to a demodulator which returns the cantilever frequency shift \emph{versus} time.
	(g) As a result of the on-off microwave modulation, the spin-induced cantilever frequency shift oscillates at a frequency of $f_{\mathrm{mod}} = 1/(2 \, t_{\mathrm{mod}}) = \SI{51}{\hertz}$.
(h) The time-dependent cantilever frequency shift, shown in low-pass filtered form (bandwidth = $\SI{250}{\hertz}$), is sent to a lock-in amplifier (not shown) to obtain the spin-induced frequency shift $\Delta f_{\mathrm{c}}$.} 
\label{fig:modulation}
\end{figure}

The sample's electron spin magnetization interacts with the second derivative of the tip's magnetic field to create a spin-force gradient that shifts the spring constant ($\Delta k$) and mechanical resonance frequency ($\Delta f_{\mathrm{c}}$) of the cantilever \cite{Garner2004jun,Moore2009dec}.
To create a distinguishable spin-induced frequency shift and to minimize heating,  the resonant microwaves are modulated in an on-off pattern synchronized with the cantilever oscillation \cite{Moore2009dec}.
This modulation pattern yields a spin-induced $\Delta f_{\mathrm{c}}$ oscillating faster than the slow $1/f$ fluctuations in cantilever frequency typically present near a sample surface \cite{Yazdanian2008jun,Yazdanian2009jun,Isaac2016apr,Isaac2017jun,Isaac2018jan}.
The resulting time-dependent $\Delta f_{\mathrm{c}}$ is observed with a frequency demodulator and a lock-in amplifier. 
See Figure~\ref{fig:modulation}.

In the spin-modulation and detection protocol of Fig.~\ref{fig:modulation} the microwave irradiation is applied in a saturating burst; we show two possible burst patterns.   
In the Fig.~\ref{fig:modulation}b experiment, employed by Moore in coworkers in Ref.~\citenum{Moore2009dec}, the resonant microwave irradiation is applied for a half cantilever cycle ($\SI{77}{\micro\second}$). 
During this half cycle the moving resonant slice sweeps out an extended region of saturated electron-spin maximization in the sample, giving rise to a large total signal.
In the Fig.~\ref{fig:modulation}c experiment, in contrast, the microwave irradiation is applied for only a short time ($\leq \SI{1}{\micro\second}$).
While the resulting per-spin signal is somewhat smaller, the resonant slice is now essentially stationary relative to the sample spins and consequently the spatial resolution in an imaging experiment is dramatically improved.

\begin{figure}
\includegraphics
	[width=\widthofonecolumn]
	{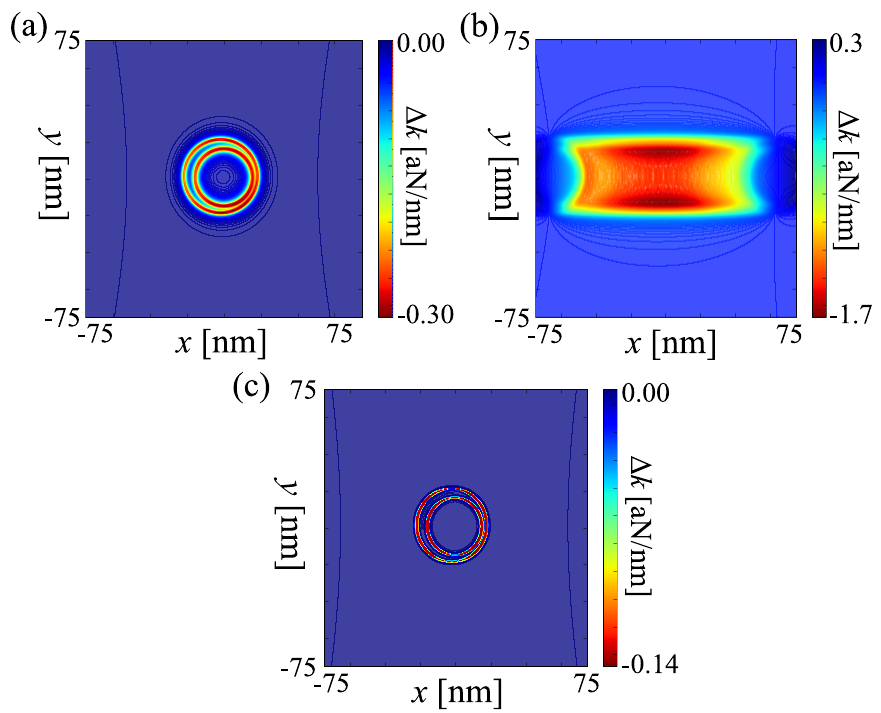}
\caption{Simulated force-gradient signal maps.
	(a) Stationary cantilever and continuous microwave irradation.
	Moving cantilever with a 
	(b) long microwave burst, Fig.~\ref{fig:modulation}b, and 
	(c) a short microwave burst, Fig.~\ref{fig:modulation}c.  
	Simulation parameters: rotating-frame microwave field intensity $B_1 = \SI{50}{\micro\tesla}$ and (b,c)  inter-burst delay $n = 1$ and cantilever zero-to-peak amplitude $x_{0\mathrm{p}} = \SI{63.5}{\nano\meter}$.}
\label{fig:compare}
\end{figure}

The expected dramatic improvement in spatial resolution can be seen clearly in the imaging simulations of Fig.~\ref{fig:compare} (discussed in detail below).
As a point of comparison, in Fig.~\ref{fig:compare}a we show the map of force-gradient signal \emph{vs}.\ cantilever $(x,y)$ position expected when the cantilever zero-to-peak oscillation amplitude has been set to zero, $x_{0\mathrm{p}} = 0$.
Distinct signal rings are seen as the resonant slice intersects each electron spin in the sample.
The associated experiment would have a terrible signal-to-noise ratio, however, since the uncertainty in the measured cantilever frequency is large when $x_{0\mathrm{p}} = 0$ \cite{Yazdanian2008jun}.
Kuehn and coworkers showed that there is an optimal cantilever amplitude which maximizes the signal-to-noise ratio in a single spin force-gradient experiment \cite{Kuehn2008feb}.
This maximum arises as a compromise between frequency noise decreasing $\propto 1/x_{0\mathrm{p}}$ and the spin signal falling off at large $x_{0\mathrm{p}}$; as $x_{0\mathrm{p}}$ increases, the spin spends an increasing fraction of the oscillation period at a large lateral distance from the tip.
Setting the amplitude to the Kuehn optimum, $x_{0\mathrm{p}} = 0.47 (r_{\mathrm{tip}} + h) = \SI{63.5}{\nano\meter}$ here, we simulated signal map for the Fig.~\ref{fig:modulation}b experiment.
In the resulting signal map, Fig.~\ref{fig:compare}b, the individual-electron signal rings of Fig.~\ref{fig:compare}a have been blurred by the cantilever oscillation.
This blurring effect renders the two electron spins unresolveable.
The simulated signal map for the  Fig.~\ref{fig:modulation}c experiment with $x_{0\mathrm{p}} = \SI{63.5}{\nano\meter}$, in contrast, shows a sharp ring of signal from each electron in the sample.
Each spin's $(x,y)$ coordinate corresponds to the center of a signal ring, and the spin's $z$ coordinate can be inferred from the radius of the signal ring.

\section{Numerical Simulation of Scanned Probe Microscope Signal}
\label{sec:simulation}

\begin{table}
\begin{tabular}{rl}
\hline
\multicolumn{2}{ l }{\textbf{A. Sample parameters}} \\
\hline
spin-lattice relaxation time & $T_1 = \SI{1.3}{\milli\second}$ \\
spin dephasing time & $T_2 = \SI{0.45}{\micro\second}$ \\
gyromagnetic ratio & $\gamma_{\mathrm{e}} = \SI{2\pi}{\radian} \times \SI{28.0}{\mega\hertz\per\milli\tesla}$ \\
saturation field & $B_{\mathrm{sat}} = 2\pi \big/ \gamma_{\mathrm{e}}\sqrt{T_1 T_2} = \SI{1.5}{\micro\tesla}$ \\
homogeneous linewidth & $B_{\mathrm{hom}} = 2\pi \big/ \gamma_{\mathrm{e}}T_2 = \SI{80}{\micro\tesla}$ \\
\hline
\multicolumn{2}{ l }{\textbf{B. Operating conditions}} \\
\hline
temperature & $T_0 = \SI{2.1}{\kelvin}$ \\
static magnetic field & $B_{\mathrm{ext}} = \SI{1200}{\milli\tesla}$ \\
microwave magnetic field\footnote{unless otherwise stated} & $B_1 = \SI{10}{\micro\tesla}$ \\
microwave frequency & $f_{\mathrm{MW}} = \SI{39.4}{\giga\hertz}$ \\
resonance frequency & $B_{\mathrm{res}} = 2\pi f_{\mathrm{MW}} \big/ \gamma_{\mathrm{e}} = \SI{1400}{\milli\tesla}$ \\
\hline
\multicolumn{2}{ l }{\textbf{C. Cantilever parameters}} \\
\hline
resonance frequency & $f_c = \SI{6.5}{\kilo\hertz}$ \\
spring constant & $k_c = \SI{1e6}{\atto\newton\per\nano\meter}$ \\
ringdown time & $\tau_c = \SI{3}{\second}$ \\
tip magnetization & $\mu_0 M_{\mathrm{s}} = \SI{1800}{\milli\tesla}$ \\
tip radius & $r_{\mathrm{tip}} = \SI{75}{\nano\meter}$ \\
tip-spin distance & $h = \SI{60}{\nano\meter}$ \\
zero-to-peak amplitude & $x_{0\mathrm{p}} = \SI{63.5}{\nano\meter}$ \\
\hline
\hline
\end{tabular}
\caption{Simulation parameters.}
\label{table:sim_param}
\end{table}

\noindent{\bf Equilibrium magnetization} --- In a magnetic field $\bm{B}_0 = B_0 \, \hat{z}$ at temperature $T_0$, the average electron spin polarization is
\begin{equation}
\label{eqn:pol_therm}
p_{\mathrm{e}} = \tanh\left(\frac{\hbar\gamma_{\mathrm{e}} B_0}{2k_B T_0}\right)
\end{equation}
with $\hbar$ the reduced Planck's constant, $k_B$ Boltzmann's constant, and $\gamma_{\mathrm{e}} = 2\pi\times 28.0$ \si{\giga\hertz\per\tesla} the electron  gyromagnetic ratio. 
Here we assume $T_0 = 2.1$ \si{\kelvin} and $B_0 = 1.4$ \si{\tesla}; under these conditions the electron spin polarization is $p_{\mathrm{e}} = 0.42$.
In our experiment the sample spins experience a magnetic field from both the external magnet and the cantilever's magnetic tip, $\bm{B_0} = \bm{B}_{\mathrm{ext}} + \bm{B}^{\text{tip}}$.
Because $\bm{B}_{\mathrm{ext}} = B_{\mathrm{ext}} \, \hat{z}$ and $B_{\mathrm{ext}} \gg B^{\mathrm{tip}}$, to first order $B_0 \approx B_{\text{ext}} + B^{\mathrm{tip}}_{z}$.
The equilibrium magnetization is $\mu_z^{\mathrm{eq}} = p_{\mathrm{e}} \mu_{\mathrm{e}}$ with $\mu_{\mathrm{e}}$ the electron magnetic moment.

\noindent {\bf Target molecule} --- 
We simulated the imaging of electron spins in a derivative of the T4 lysozyme molecule: mutant K65V1/R76V1 in which a nitroxide spin probe is attached to each of two cysteine residues by a disulfide linkage \cite{Langen2000jul, Jacobsen2005jun, Jacobsen2006jun, ToledoWarshaviak2010oct} (Fig.~\ref{fig:3K2R-PSF} inset). 
The crystal structure of this molecule was obtained from the protein database (pdb 3K2R \cite{ToledoWarshaviak2010oct}).
Coordinates were extracted from the pdb file, converted from units of \AA\ to units of \si{\nano\meter}, and translated by the vector $(+2, 0, -5) \: \si{\nano\meter}$. 
The coordinates of the nitroxides' oxygen atoms were extracted \textit{via} Chimera \cite{Pettersen2004oct} and used as the electron-spin coordinates.
The protein was placed at a vertical distance of $h = \SI{60}{\nano\meter}$ below the surface of the spherical tip to mimic an experiment with a $\SI{20}{\nano\meter}$ tip-sample separation, a $\SI{10}{\nano\meter}$ metal coating, a $\SI{20}{\nano\meter}$ spin ``dead layer'', with the spin located an additional $\SI{10}{\nano\meter}$ below the dead layer. 
Experimentally the tip-sample separation is chosen as a compromise between large force-gradient signal and large surface-induced frequency noise at close tip-sample separation.

\noindent {\bf Magnetic resonance parameters} --- 
The electron magnetic resonance parameters are given in Table ~\ref{table:sim_param}A.
These parameters are drawn from the experiments of Ref.~\onlinecite{Moore2009dec} performed at $T_0 = \SI{4.2}{\kelvin}$ and $B_0 = \SI{0.6}{\tesla}$. 
To saturate nitroxide electron spins requires a transverse field amplitude $B_1$ larger than $B_{\mathrm{sat}} = 2\pi/\gamma_{\mathrm{e}}\sqrt{T_1 T_2} = \SI{1.5}{\micro\tesla}$. 
We assume $B_1 = \SI{10}{\micro\tesla}$, except where noted (Fig.~\ref{fig:compare}).
The microwave frequency, $f_{\mathrm{MW}} = \SI{39.4}{\giga\hertz}$, was chosen to bring spins into resonance at a distance $h = \SI{60}{\nano\meter}$ directly below the spherical tip.

In a scanned force-gradient signal map, the sharpness of observed signal rings in our detection protocol is simply related to the length of the pulse $\tau_{\mathrm{p}}$. 
While the pulse is active, the cantilever sweeps a finite distance $\Delta x$ and thus drags the sensitive slice along, blurring the signal map by $\Delta x$.
Hence, the maximum length of the microwave burst $\tau_{\mathrm{p}}$ depends on the acceptable blurring $\Delta x$ of the resonant slice and the velocity of the cantilever tip $v_c$, and is given by
\begin{equation}
\label{eqn:tau_p}
\tau_{\mathrm{p}} 
  = \dfrac{\Delta x}
          {v_c^{\mathrm{max}}} 
  = \frac{\Delta x}
         {2\pi f_c x_{0\mathrm{p}}}
\end{equation}
with $f_c$ the cantilever's resonance frequency and $x_{0\mathrm{p}}$ the zero-to-peak amplitude of the cantilever oscillation. 
The velocity $v_c$ of the cantilever is calculated by writing $x_c(t) = x_{0\mathrm{p}} \sin{\left(2 \pi f_{\mathrm{c}} t \right)}$ and taking the time derivative of $x_c$.
The velocity reaches a maximum value $v_c^{\mathrm{max}} = 2\pi f_c x_{0\mathrm{p}}$ at times when $x_c = 0$. 
For a typical cantilever resonance frequency of $f_c = \SI{6.5}{\kilo\hertz}$ and oscillation amplitude of $x_{0\mathrm{p}} = \SI{63.5}{\nano\meter}$, the maximum velocity is $v_c^{\mathrm{max}} = \SI{2.59}{\nano\meter/\micro\second}$. 
At this velocity, the time required for the cantilever to sweep $1$ \si{\nano\meter} at its maximum velocity is $0.4$ \si{\micro\second}. 
Thus, to keep the blurring to $\Delta x \leq 2$ \si{\nano\meter} we need to keep the pulse time $\tau_{\mathrm{p}}\lesssim 0.8$ \si{\micro\second}.

\noindent {\bf Magnetic resonance} --- 
Now consider the effect of the cyclic microwave irradiation on the individual electron spins in the sample.
In the presence of an oscillating microwave magnetic field, only those spins inside the resonant slice satisfy the magnetic resonance condition, have their magnetization modulated, and contribute to signal.
The resonant slice is defined as the set of points $\bm{r}$ in the sample where 
\begin{equation}
\label{eqn:B_res}
 B_{\mathrm{ext}} + B_z^{\mathrm{tip}}(\bm{r}) 
 = 2 \pi 
   f_{\mathrm{MW}} / \gamma_{\mathrm{e}}
\end{equation}
with $f_{\mathrm{MW}}$ the frequency of the the applied microwave field.
In our coordinate system, location $x = 0$ and $y = 0$ are chosen to be the center of the scanned signal map. 
The location of $z = 0$ is at a vector ($0,0,+60$) \si{\nano\meter} relative to the center of the spherical magnetic tip.   
Thus $\bm{r} = (0,0,0)$ is defined to be the nadir point of the sensitive slice when the tip is located at the center of the scanned signal map.
In the Moore experiment of Ref.~\citenum{Moore2009dec}, the microwave-induced change in magnetization was taken to be $\Delta \mu_z = \mu_z^{\mathrm{ss}} - \mu_z^{\mathrm{eq}}$ with $\mu_z^{\mathrm{ss}}$ the magnetization obtained from the steady-state solution to the Bloch equations:
\begin{equation}
\label{eqn:delta_mu_ss_long}
\Delta \mu_z 
  = - \frac{
    p_{\mathrm{e}} 
    \mu_{\mathrm{e}} 
    \gamma_{\mathrm{e}}^2 B_1^2 T_1 T_2
  }{
    1 + T_2^2 \Delta \omega_0^2 
    + \gamma_{\mathrm{e}}^2 B_1^2 T_1 T_2
  }
\end{equation} 
with $B_1$ the amplitude of the applied microwave field, $T_1$ the electron spin-lattice relaxation time, $T_2$ the electron spin dephasing time, and 
\begin{equation}
\Delta \omega_0 
  = \gamma_{\mathrm{e}} B_{\mathrm{ext}}
  + \gamma_{\mathrm{e}} 
      B_z^{\mathrm{tip}}(\bm{r})
-  2\pi f_{\mathrm{MW}}
\end{equation}
the resonance offset.  
The result in Eq.~\ref{eqn:delta_mu_ss_long} is not strictly applicable in the Fig.~\ref{fig:modulation}b experiment since the microwave field is applied intermittently.
For simplicity we nevertheless use the change in magnetization from Eq.~\ref{eqn:delta_mu_ss_long} as an approximation to simulate the signal map for the Fig.~\ref{fig:modulation}b experiment.

In the case of the Fig.~\ref{fig:modulation}c experiment, the microwave burst is very short relative to the spin-lattice relaxation time $T_1$. 
As a result of the microwaves being applied in intermittent bursts, the spins do not attain the steady state magnetization given by Eq.~\ref{eqn:delta_mu_ss_long}; they instead converge to a different steady-state magnetization which we wish to calculate.
Using Torrey's result,\cite{Torrey1949oct} the magnetization due to the short microwave irradiation for on-resonance spins, in the limit $T_2\ll T_1$, is given by
\begin{equation}
\label{eqn:postpulse-muz}
\mu_z(0^+) = \mu_z(0^-) \mathcal{L}(\tau_{\mathrm{p}}) + \mu_z^{\mathrm{eq}} \mathcal{S}(\tau_{
\mathrm{p}}),
\end{equation}
with $\mu_z(0^-)$ and $\mu_z(0^+)$ the magnetization before and after the microwave burst, respectively, and
\begin{align}
\mathcal{L}(\tau_{\mathrm{p}})
	&= E_2 
	\cos\left(\frac{\tau_{\mathrm{p}} \kappa}{2 T_2}\right) 
	+ \frac{E_2}{\kappa}
			\sin\left(\frac{\tau_{\mathrm{p}} \kappa}{2 T_2}\right)
	\label{eq:intermittant-Bloch-L} \\
\mathcal{S}(\tau_{\mathrm{p}}) 
	&= D[1 - \mathcal{L}(\tau_{\mathrm{p}})] 
		+ \frac{2T_2}{T_1}\frac{E_2}{\kappa}\sin\left(\frac{\tau_{\mathrm{p}}\kappa}{2 T_2}\right).
	\label{eq:intermittant-Bloch-S}
\end{align}
In Eqs.~\ref{eq:intermittant-Bloch-L} and \ref{eq:intermittant-Bloch-S}, $\kappa = \sqrt{4\gamma_{\mathrm{e}}^2 B_1^2 T_2^2 -1 }$ is the unitless frequency of oscillation during irradiation, $E_2 = e^{-\tau_{\mathrm{p}}/(2 T_2)}$ is a factor accounting for magnetization relaxation during irradiation, and $D = 1/(1+\gamma_{\mathrm{e}}^2\, T_1\, T_2 )$ is the steady-state $z$-magnetization for on-resonance irradiation from the Bloch equations.
In the Fig.~\ref{fig:modulation}c experiment the microwave burst is followed by a period of microwave-free relaxation lasting a duration $\tau = n \, T_c$, with $T_c$ the cantilever period and $n$ an integer.
During this relaxation period the electron-spin magnetization $\mu_z(0^+)$ recovers towards its equilibrium value $\mu_z^{\mathrm{eq}}$.
Using the Bloch equation and the initial condition --- the magnetization at time $t = 0$ is $\mu_z(0^{+})$ --- the magnetization during the relaxation period is
\begin{equation}
\label{eqn:relaxing-muz}
\mu_z(t) 
  = \mu^{\mathrm{eq}} 
    + [\mu_z(0^{+}) - \mu_z^{\mathrm{eq}}] 
    \: e^{-t/T_1} .
\end{equation}
The $z$ component of the electron-spin magnetization before nutation and at the end of the relaxation period should be equal at steady-state,
\begin{equation}
\label{eqn:steady-state-condition}
\mu_z(\tau) = \mu_z(0^{-}). 
\end{equation}
Inserting Eq.~\ref{eqn:steady-state-condition} into Eq.~\ref{eqn:postpulse-muz}, inserting the resulting equation into Eq.~\ref{eqn:relaxing-muz} evaluated at time $t = \tau$, and solving for $\mu_z(0^{-})$ gives
\begin{equation}
\label{eqn:ss-muz0-}  
\mu_z(0^{-}) 
 = \mu_z^{\mathrm{eq}} \, 
 \frac{1 - E_1 - E_1 \: \mathcal{S}(\tau_{\mathrm{p}})}
 {1 - E_1 \: \mathcal{L}(\tau_{\mathrm{p}})}
\end{equation}
with
\begin{equation}
\label{eq:E1-defn}
E_1 = e^{-\tau/T_1}
\end{equation}
a factor accounting for magnetization relaxation in between microwave bursts.
Inserting Eq.~\ref{eqn:ss-muz0-} into Eq.~\ref{eqn:relaxing-muz} and solving for the \emph{change} in magnetization $\Delta \mu_z(t) = \mu_{z}(t) - \mu_z^{\mathrm{eq}}$ induced by microwave irradiation gives
\begin{equation}
\label{eqn:ss-muz(t)-norm}
\Delta \mu_z(t) = \mu^{\mathrm{eq}} \:
	\frac{\mathcal{L}(\tau_{\mathrm{p}}) + \mathcal{S}(\tau_{\mathrm{p}}) -1}
				{1-E_1 \:\mathcal{L}(\tau_{\mathrm{p}})} \: e^{-t/T_1}.
\end{equation}
To obtain the average magnetization observed by the lock-in detector we average Eq.~\ref{eqn:ss-muz(t)-norm} over the observation period $\tau$, giving
\begin{multline}
\label{eqn:delta-muz-onres}
\langle \Delta \mu_z \rangle
 = \frac{1}{\tau} \int_{0}^{\tau} \mu_z(t) \, dt \\
 = \mu_z^{\mathrm{eq}} \frac{T_1}{\tau} (1 - E_1) 
\frac{\mathcal{L}(\tau_{\mathrm{p}}) + \mathcal{S}(\tau_{\mathrm{p}}) -1}
				{1-E_1 \: \mathcal{L}(\tau_{\mathrm{p}})}.
\end{multline}

\begin{figure}
\includegraphics
	[width = \widthofonecolumn]
	{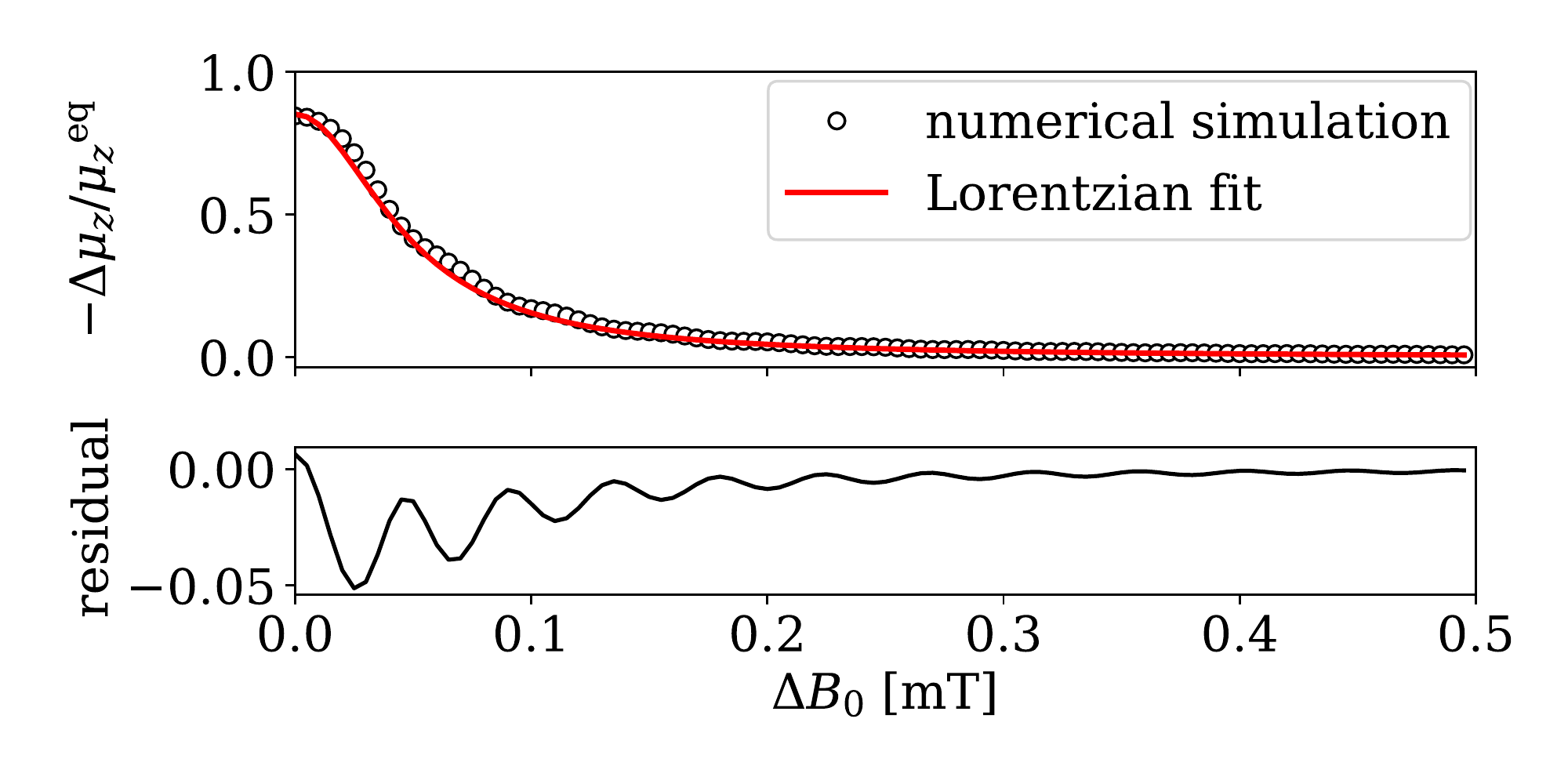}
\caption[The response of an electron spin from Bloch equation]{The numerically calculated response of an electron spin under the Fig.~\ref{fig:modulation}c modulation versus resonance offset. 
	(top) The negative change in magnetization $\Delta \mu_z$ relative to the initial equilibrium magnetization $\mu_z^{\mathrm{eq}}$ as a function of resonance offset $\Delta B_0$. The numerical result (circle) is fitted to a Lorentzian lineshape (solid line). 
	(bottom) The residuals between the fitted Lorentzian lineshape and the numerical result. Simulation parameters: $\tau_{\mathrm{p}} = 0.8$ \si{\micro\second} and $\tau = T_c = 154$ \si{\micro\second}.}
\label{fig:Bloch_numerical}
\end{figure}

The change in magnetization expressed in Eq.~\ref{eqn:delta-muz-onres} is only valid for on-resonance spins. 
An analytic expression similar to Eq.~\ref{eqn:delta-muz-onres} for spins at a resonant offset $\Delta B_0 = \Delta \omega_0/\gamma_{\mathrm{e}}$ is hard to obtain. 
We instead numerically simulated the Bloch equations and obtained the detected average magnetization for the protocol in Fig.~\ref{fig:modulation}c as a function of the resonance offset $\Delta B_0$. 
The steady-state response of an electron spin is shown in Fig.~\ref{fig:Bloch_numerical}.
The $\mu_z$ \emph{vs.} $\Delta B_0$ lineshape is well described by a Lorentzian of the form
\begin{equation} 
\label{eqn:response-lineshape}
\Delta \mu_z (\Delta B_0) = \Delta \mu_z (0) \frac{b^2}{\Delta B_0^2 + b^2}
\end{equation}
where $\Delta \mu(0)$ is the on-resonance response given by Eq.~\ref{eqn:delta-muz-onres} and $b$ is the half-linewidth of the response function.
We find empirically that $b$ is well approximated as
\begin{equation}
\label{eqn:halfwidth}
b = \frac{B_{\mathrm{hom}}}{2} \sqrt{\frac{B_1^2}{B_{\mathrm{hom}} B_{\mathrm{sat}}}+\frac{T_2}{\tau_{\mathrm{p}}}}
\end{equation}
with $\tau_{\mathrm{p}}$ the length of the microwave burst, $B_{\mathrm{hom}} = 2\pi/(\gamma_{\mathrm{e}} T_2)$, and  
$B_{\mathrm{sat}} = 2\pi/(\gamma_{\mathrm{e}} \sqrt{T_1 T_2})$.

From Eq.~\ref{eqn:halfwidth} and the simulation parameters in Table~\ref{table:sim_param} we can estimate the full linewidth of the cyclic-saturation experiment with intermittent irradiation to be $\Delta B = 2b = \SI{94}{\micro\tesla}$.
Using a lateral gradient of $G_x = \SI{5.5}{\milli\tesla/\nano\meter}$ from Ref.\citenum{Longenecker2012nov}, we can estimate the linewidth-limited resolution to be $\Delta B/G_x = \SI{0.17}{\angstrom}$.
Equation~\ref{eqn:response-lineshape} is a Bloch-like equation that describes the steady state achieved in the intermittent-pulse experiment of Fig.~\ref{fig:modulation}c for spins at various resonance offsets.
We therefore use Eq.~\ref{eqn:response-lineshape} to simulate the scanning signal following the protocol in Fig.~\ref{fig:modulation}c, as well as to reconstruct the image from the measured signal.

\noindent{\bf Tip model} ---
In this simulation we used a simplified spherical-tip model. 
The tip radius, $r_{\mathrm{tip}} = \SI{75}{\nano\meter}$, was chosen to yield a field gradient similar to that of the nanorod magnet prepared by Longenecker and coworkers \cite{Longenecker2012nov}.
The $z$-component of the magnetic field at the sample location $\bm{s} = (x,y,z)$ relative to the center of a spherical magnet located at ($0,0,0$) is given by \cite{Kempf2003feb, Kuehn2008feb}
\begin{equation}
\label{eqn:Bz_tip}
B_z^{\mathrm{tip}} (x,y,z) 
= \frac{\mu_0 M_s}{3} 
  r_{\mathrm{tip}}^3 
  \frac{2z^2 - x^2 - y^2}
       {(x^2 + y^2 + z^2)^{5/2}}
\end{equation}
where $\mu_0 M_s = 1.8$ \si{\tesla} is the tip magnetization of a cobalt magnet. The tip field's lateral gradient $G_x$ was calculated by taking the partial derivative of the $z$-component of the tip field $B_z^{\mathrm{tip}}$ in Eq.~\ref{eqn:Bz_tip}.
The result is
\begin{align}
G_x (x,y,z) &= \frac{\partial B_z^{\mathrm{tip}} (x,y,z)}{\partial x} \nonumber \\
\label{eqn:Gx}
&= \mu_0 M_s r_{\mathrm{tip}}^3 \:x\: \frac{x^2+y^2-4z^2}{(x^2+y^2+z^2)^{7/2}}.
\end{align}

\noindent {\bf Cantilever amplitude} --- The cantilever zero-to-peak amplitude was set to the optimal value for a single-spin frequency-detected magnetic resonance experiment \cite{Kuehn2008feb}, $x_{0\mathrm{p}} = 0.47 (r_{\mathrm{tip}} + h) \approx \SI{63.5}{\nano\meter}$.

\noindent {\bf Scanning} --- To obtain a two-dimensional force-gradient signal map, the cantilever is raster-scanned in the $(x,y)$ plane to collect a force-gradient map. 
Here we simulate a 2D scanning grid of 128 $\times$ 128 pixels covering an area of $50$ \si{\nano\meter} $\times$ $50$ \si{\nano\meter}.

\noindent {\bf Force-gradient signal} ---
At every location of the cantilever $\bm{r} = (x,y,z)$, we simulate the tip oscillating over the sample around the cantilever location, sweeping from $(x-x_{0\mathrm{p}},y,z)$ to $(x+x_{0\mathrm{p}},y,z)$. 
The lateral position of the magnetic tip at time $t$ relative to the zero crossing position of the cantilever is given by
\begin{equation}
\label{eqn:x_cant}
x_c(t) = x_{0\mathrm{p}} \sin(2\pi f_c t) = x_{0\mathrm{p}} \sin\vartheta
\end{equation}
where $x_{0\mathrm{p}}$ is the zero-to-peak amplitude of the cantilever oscillation, $f_c$ is the cantilever's resonance frequency, and $\vartheta = 2\pi f_c t$ is the phase of the cantilever at time $t$.
When the microwave bursts are cyclically applied to selectively saturate electron spins in the sample, the change in the $z$-component of the spin magnetic moment $\Delta \mu_z$ causes a shift in the cantilever force constant $\Delta k$ through its interaction with the magnetic tip. As shown by Lee \textit{et al.}, \cite{Lee2012apra} the resulting shift in the spring constant of the cantilever can be calculated as
\begin{equation}
\label{eqn:delta_k_full}
\Delta k = \frac{1}{\pi x_{0\mathrm{p}}} \sum_{j=1}^{N_s} \int_{-\pi}^{\pi} \Delta \mu_z(\bm{s}_j - \bm{r}(\vartheta)) G_x(\bm{s}_j - \bm{r}(\vartheta))  \cos{\vartheta}\: d\vartheta
\end{equation} 
where the index $j$ implements a sum over the $N_s$ spins in the sample;
 $\Delta \mu_z(\bm{s}_j - \bm{r}(\vartheta)) = \Delta \mu_z(\Delta B_0(\bm{s}_j - \bm{r}(\vartheta))$, Eq.~\ref{eqn:response-lineshape}, is the change in the $z$-component of the magnetic moment of the electron spin $j$ at location $\bm{s}_j = (x_j,y_j,z_j)$ relative to the tip location $\bm{r}(\vartheta) = (x + x_{0\mathrm{p}}\sin\vartheta,y,z)$ corresponding to the cantilever phase $\vartheta$;  
and $G_x$, Eq.~\ref{eqn:Gx}, is the lateral magnetic field gradient at location $\bm{s}_j$ of spin $j$, relative to the location $\bm{r}(\vartheta)$ of the center of the spherical magnetic tip.
The spring constant shift signal $\Delta k$ was numerically calculated by dividing the cantilever oscillation into $N_t = 65$ points, with an associated time step of $\Delta t = 1/f_c(N_t - 1)$, and calculating the phase integral in Eq.~\ref{eqn:delta_k_full} as a sum using the trapezoid rule. 
At each time point $t_q = q\Delta t$ corresponding to cantilever phase $\vartheta_q = 2\pi q/(N_t -1)$, with $q = 0,1,... N_t-1$, the force acting on the cantilever from all the spins in the sample was the sum of the time-dependent force contributed by each individual spin.
Thus, we rewrite Eq.~\ref{eqn:delta_k_full} as a discrete sum over the cantilever period
\begin{equation}
\label{eqn:delta_k_disc}
\Delta k =  \frac{2\pi f_c \Delta t}{\pi x_{0\mathrm{p}}^2} \sum_{j=1}^{N_s} \sum_{q=0}^{N_t-1} \Delta \mu_z (\bm{s}_j - \bm{r}(t_q)) G_x(\bm{s}_j - \bm{r}(t_q)) x_c(t_q)
\end{equation}
where and $\bm{r}(t_q) = (x + x_c(t_q), y, z)$ is the location of the tip at time $t_q$. 
The change in the magnetic moment $\Delta \mu_z$ of spin $j$, relative to the equilibrium magnetization $\mu_{\mathrm{e}}$, is dependent on the relative location of the spin and the tip, and calculated by plugging into Eq.~\ref{eqn:response-lineshape} 
the on-resonance response $\Delta\mu_z(0) = \expval{\Delta\mu_z}$ from Eq.~\ref{eqn:delta-muz-onres}, the expected half linewidth $b$ of Eq.~\ref{eqn:halfwidth}, and the resonance offset $\Delta B_0(\bm{s}_j - \bm{r}(t_q)) = B_{\mathrm{ext}} - B_z^{\mathrm{tip}}(\bm{s}_j - \bm{r}(t_q)) - 2 \pi f_{\mathrm{MW}} / \gamma_{\mathrm{e}}$ with $B_z^{\mathrm{tip}}(\bm{s}_j - \bm{r}(t_q))$ given by Eq.~\ref{eqn:Bz_tip}.

\noindent {\bf Noise} --- We added Gaussian white noise to the simulated force-gradient maps to demonstrate the reconstruction algorithm's ability to obtain individual electron coordinates from realistic, noisy data.
The cantilever's force-gradient thermal noise floor is given by \cite{Albrecht1991jan, Yazdanian2008jun}
\begin{equation}
P_{\delta k}^{\mathrm{therm}} 
  = \frac{2 k_B T_0 \, k_c}
         {\pi^2 \, f_c^2 \, \tau_c \,  x_{0\mathrm{p}}^2}
\end{equation}
with $k_B$ Boltzmann's constant; $k_c$, $\tau_c$, $f_c$, and $x_{0\mathrm{p}}$ given in Table~\ref{table:sim_param}B; and $T_0$ defined in Table~\ref{table:sim_param}B.
Using representative values given in Table~\ref{table:sim_param}(A,B) we estimate the thermal noise at $T_0 = 2.1$ \si{\kelvin} to be $P_{\delta k}^{\mathrm{therm}} = \SI{5.75e3}{\atto\newton^2\per\nano\meter^2 \hertz}$.
The expected variance of the force-gradient noise is given by
\begin{equation}
\label{eqn:noise-var}
\sigma_{\delta k}^2 
  = \frac{P_{\delta k}^{\mathrm{therm}}}{T_{\mathrm{avg}}}
\end{equation}
with $T_{\mathrm{avg}}$ the per-data-point averaging time.
For an averaging time of $T_\mathrm{avg} = \SI{3}{\second}$ per data point, the expected root-mean-square (rms) of the thermal force-gradient noise is $\sigma_{\delta k} \approx \SI{0.06}{\atto\newton\per\nano\meter}$.
This rms force-gradient noise is equivalent to a force noise of $\delta F_{\mathrm{rms}} = \sigma_{\delta k} x_{0\mathrm{p}}/\sqrt{2} = \SI{2.7}{\atto\newton}$. 
Gaussian random noise with a mean of zero and a variance of $\sigma_{\delta k}$, ${\cal N}(0,\sigma_{\delta k})$, was added to the calculated force-gradient signal at $(x,y)$ to obtain the simulated force-gradient signal map in Fig.~\ref{fig:reconst-Bayes}a.
Additional noise caused by the random fluctuations of the electron spins is neglected since we are working in the limit where detector noise is the dominant noise source.
See Appendix~\ref{sec:supplementary2} for a justification of this assumption.

\section{Reverse Monte Carlo Reconstruction of Spin Coordinates}
\label{sec:Bayesian}

\begin{figure}
\includegraphics[width=\widthofonecolumn]{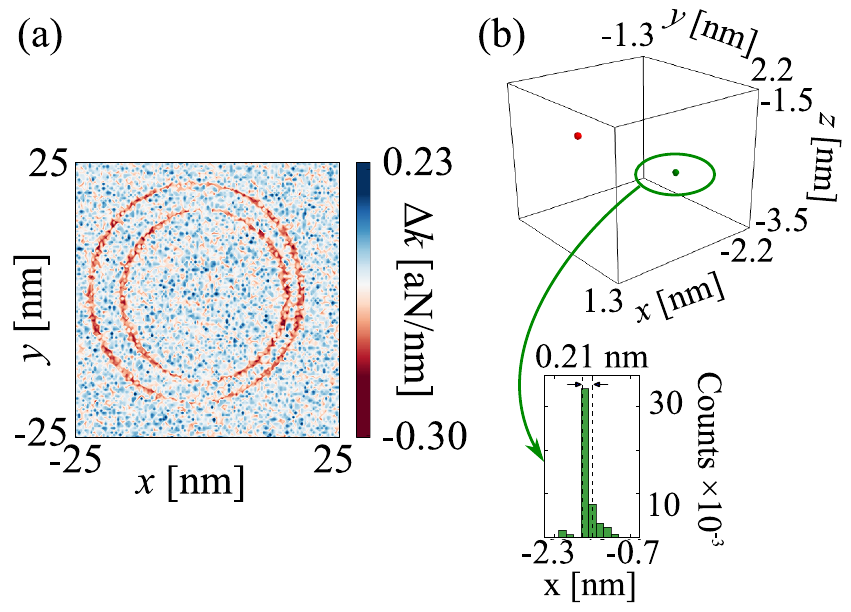}
\caption{
	(a) Simulated 2D scanned force-gradient signal for the doubly spin-labeled T4 lysozyme mutant pdb 3K2R.
	(b) Image reconstructed from the simulated signal using the Bayesian Markov-chain Monte-Carlo approach discussed in the text. 
	(Upper) Reconstructed three-dimensional spin density showing the location of two individual electron spins separated by $\SI{21}{\angstrom}$.
	(Lower) The posterior distribution of the $x$-position of one of the two electron spins, showing a resolution of $\sim \SI{2.1}{\angstrom}$.
	Simulation parameters: 
$p_{\mathrm{e}} = 0.42$ and 
$\tau_{\mathrm{p}} = \SI{0.8}{\micro\second}$. 
	Reconstruction parameters:  
step size $\sigma_{\delta\bm{R}} = \SI{1}{\nano\meter}$ and number of iterations $= 5\times 10^4$.
}
\label{fig:reconst-Bayes}
\end{figure}

In order to obtain an image of the electron spins in the sample, we fit the simulated signal of Fig.~\ref{fig:reconst-Bayes}(a) to  Eq.~\ref{eqn:delta_k_disc} with the electron spins' coordinates as the fitting parameters. 
This fitting is done using a reverse Markov-chain Monte Carlo approach in which the electron-spin coordinates are randomly varied and the variations accepted with a probability taken from Bayesian analysis.
Once the electron coordinates have converged, the error bars of the spins' coordinates are determined from the distribution of the spins' coordinates around their equilibrium positions. 

In this section, for mnemonic simplicity, the measured signal will be called ${S}$;
the calculated signal, $\Delta k$ in Eq.~\ref{eqn:delta_k_disc}, will be written as $\hat{S}(\bm{R})$ to emphasize its dependence on the spin  coordinates $\bm{R} = (x_1, y_1, z_1, x_2, y_2, z_2)$; 
and the noise variance, $\sigma_{\delta k}^2$ in Eq.~\ref{eqn:noise-var}, will be referred to as simply $\sigma^2$.
The conditional probability $\mathcal{P}(\bm{R}|{S},\sigma^2)$ for the two electron spins to have a specific set of coordinates $\bm{R}$ is calculated as
\begin{multline}
\label{eqn:prior_prob}
\mathcal{P} (\bm{R}|{S},\sigma^2) = \frac{1}{(2\pi\sigma^2)^{N/2}} \\
\times \exp\left(-\frac{1}{2 \sigma^2}\sum_{i=0}^{N-1}\left|\hat{S}_i(\bm{R})-S_i\right|^2\right)
\end{multline}
where the sum implemented by the index $i$ is over $N$ grid points in the signal map; and
$S_i$ and $\hat{S}_i$ are the measured and calculated signal, respectively, at location $\bm{r}_i$.
For simplicity, we further assume that the only source of noise in the measured signal is the thermal noise $\sigma^2$ which we take to be the same at every grid point. 
The probability $\mathcal{P}(\bm{R}|{S},\sigma^2)$ is normalized over all realizations of the error $\delta S_i = \hat{S}_i(\bm{R}) - S_i$ at all pixels $i$ in the measured signal map.

At every iteration $n$, we generate a new set of coordinates
\begin{equation}
\bm{R}_{\mathrm{new}} 
  = \bm{R}_{n-1} + \delta \bm{R}
\end{equation}
from the previous coordinates $\bm{R}_{n-1}$, by applying a perturbation $\delta\bm{R} \sim {\cal N}(0,\sigma_{\delta \bm{R}}^2)$. 
By changing the distribution's variance $\sigma_{\delta \bm{R}}^2$, for example, we can adjust the burn-in time, the period during which the fitting parameters $\bm{R}$ wander around before reaching equilibrium.
Although there is no set requirement for burn-in time, we would like to keep it less than 30\% of the total calculation time.
The optimal acceptance rate is thought to be 23.4\% for a Markov chain of infinite dimension.\cite{Roberts1997feb, Sherlock2009aug}
Here the dimensionality of our spin coordinate set $\bm{R}$ is only 6; we therefore aim for a more general acceptance rate of less than 50\% to balance the convergence of $\bm{R}$ towards equilibrium and the burn-in time.
The distribution from which we sample $\delta \bm{R}$ is another knob we can turn to optimize the Markov chain.
Since the noise in our signal is Gaussian, let us use a normal distribution $\mathcal{N}(0,\sigma_{\delta\bm{R}}^2)$ to sample $\delta \bm{R}$.
With each set of coordinates generated $\bm{R}_{\mathrm{new}}$, we calculate the force-gradient signal $\hat{S}(\bm{R}_{\mathrm{new}})$ using Eq.~\ref{eqn:delta_k_disc}. 
The acceptance probability $\alpha$ of a step in the Markov chain follows the Metropolis-Hastings algorithm: \cite{Metropolis1953jun, Hastings1970apr}
\begin{equation}
\label{eqn:MetroHast_prob}
\alpha = \min\left(1,\frac{\mathcal{P}(\bm{R}_{\mathrm{new}}|S,\sigma^2)}{\mathcal{P}(\bm{R}_{n-1}|S,\sigma^2)}\times \frac{\mathcal{P}(\bm{R}_{n-1}|\bm{R}_{\mathrm{new}})}{\mathcal{P}(\bm{R}_{\mathrm{new}}|\bm{R}_{n-1})}\right)
\end{equation}
where $\mathcal{P}(\bm{R}_{n-1}|\bm{R}_{\mathrm{new}})$ is the probability to obtain $\bm{R}_{n-1}$ from $\bm{R}_{\mathrm{new}}$, which is the same distribution from which we sample $\delta\bm{R}$, \textit{i.e.}\ $\mathcal{P}(\bm{R}_{n-1}|\bm{R}_{\mathrm{new}}) = \mathcal{N}(0,\sigma_{\delta \bm{R}}^2)$.
For a normal distribution, the ratio of $\mathcal{P}(\bm{R}_{n-1}|\bm{R}_{\mathrm{new}})$ to $\mathcal{P}(\bm{R}_{\mathrm{new}}|\bm{R}_{n-1})$ in Eq.~\ref{eqn:MetroHast_prob} is equal to 1. 
We can therefore simplify the expression of the acceptance probability $\alpha$ from Eq.~\ref{eqn:MetroHast_prob} to the expression in Metropolis's algorithm \cite{Metropolis1953jun}, 
\begin{equation}
\alpha = \min\left(1,\frac{\mathcal{P}(\bm{R}_{\mathrm{new}}|S,\sigma^2)}{\mathcal{P}(\bm{R}_{n-1}|S,\sigma^2)}\right)
\end{equation}
which expands to
\begin{multline}
\label{eqn:Metro_prob}
\alpha = \min\left(1,\exp\left[-\frac{1}{2\sigma^2}\sum_{i=0}^{N-1}\left( \left|\hat{S}_i(\bm{R}_{\mathrm{new}}) - S_i\right|^2 \right. \right. \right. \\
- \left. \left. \left.
\left|\hat{S}_i(\bm{R}_{n-1}) - S_i\right|^2\right)\right]\right).
\end{multline}
If $\alpha = 1$, the move from $\bm{R}_{n-1}$ to $\bm{R}_{\mathrm{new}}$ is always accepted and the guessed $\bm{R}_{\mathrm{new}}$ is recorded as a new link on the Markov chain by setting $\bm{R}_n = \bm{R}_{\mathrm{new}}$. 
However, if $\alpha < 1$, we can only accept $\bm{R}_{\mathrm{new}}$ with a probability of $\alpha$. For example, if $\alpha = 0.7$, we accept $\bm{R}_{\mathrm{new}}$ only 70\% of the time. 
If we reject $\bm{R}_{\mathrm{new}}$, then the previous position $\bm{R}_{n-1}$ will instead be recorded as the new link on the Markov chain, $\bm{R}_n = \bm{R}_{n-1}$. 

The result of the reverse Monte Carlo reconstruction from a simulated 2D scanned force-gradient signal with $\SI{0.06}{\atto\newton}\text{-rms}$ noise (Fig.~\ref{fig:reconst-Bayes}a) is shown in Fig.~\ref{fig:reconst-Bayes}b.
To demonstrate the reverse Monte Carlo reconstruction, we started with both spins at the origin ($0,0,0$).
For the reconstruction in Fig.~\ref{fig:reconst-Bayes}b, we used a uniform standard deviation of $\sigma_{\delta \bm{R}} = \SI{1}{\nano\meter}$ to step $\delta\bm{R}$. 
The value of $\sigma_{\delta \bm{R}}$ here was conservatively chosen based on the microwave irradiation time $\tau_{\mathrm{p}} = 0.8$ \si{\micro\second}, which blurs the signal map by $\sim \SI{2}{\nano\meter}$.
With these parameters, the reconstruction time for $5 \times 10^4$ iterations was \emph{ca}.\ 3 hours, with the burn-in taking $\sim$ 30\% of the total number of iterations. 
Since we used a relatively large step size, the acceptance rate was quite low --- less than $1\%$.
At each iteration, the probability was calculated using Eq.~\ref{eqn:prior_prob}.
We concluded that equilibrium was reached when the calculated probability settled near a maximum value.
The observed equilibrium distribution of the $x$-coordinate for one of the two spins is shown as the inset of  Fig.~\ref{fig:reconst-Bayes}b.
The apparent resolution in the $x$-direction is \SI{2.1}{\angstrom} .

The resolution observed in Fig.~\ref{fig:reconst-Bayes}b is smaller than the width of the stepping distribution $\sigma_{\delta\bm{R}}$.
Together with the fact that the acceptance was on the low end (<1\%) and a very sharp and uneven probability distribution was observed, this resolution indicates that the step size was indeed too large and that the stepping process spent a lot of time creating rejected proposals and was unable to fully sample the posterior distribution. 
One method to fine-tune the reverse Monte Carlo reconstruction, improve the resolution, and increase the acceptance rate is to change the standard deviation $\sigma_{\delta \bm{R}}$ of the sampling distribution.
Choosing too small a value for $\sigma_{\delta \bm{R}}$ will cause the reconstruction to take a longer time to reach equilibrium, however. 
On the other hand, picking a large value of $\sigma_{\delta \bm{R}}$ to quickly reach equilibrium would make the reconstruction unable to explore the full distribution of the spin coordinates at equilibrium.
Assuming no prior knowledge of the tip-spin distance besides the approximate center of the signal ring, we need a relatively large step size so that $\bm{R}$ can reach the equilibrium coordinates without spending too much time with incremental stepping.
We propose to use a one-step, fast, approximate reconstruction protocol to bring the spins closer to their true coordinates before starting the reverse Monte Carlo reconstruction process with an appropriately small step size.

\section{Fast reconstruction with Fourier deconvolution}
\label{sec:Tikh}

Here we present a fast reconstruction method based on the fast Fourier transform (FFT) and Tikhonov regularization which can quickly produce a three-dimensional spin-density map. 
Using the coordinates extracted from this density map, we can significantly speed up Bayesian reconstruction by producing the initial guesses for the spins' location close to their `true' coordinates.
Similar to previous deconvolution methods,\cite{Grinolds2011may, Grinolds2014mar} we need to discretize the sample grid for this reconstruction. 

\subsection{Point-spread function}

The signal $S(\bm{r})$ obtained by scanning the cantilever in the $(x,y)$-plane can be approximated as a convolution of a point-spread function (psf) $T(\bm{r} - \bm{s})$ with the spin density $\rho(\bm{s})$ where $\bm{r}$ and $\bm{s}$ are location vector of the cantilever tip and the spin, respectively:
\begin{equation}
\label{eqn:conv}
{S}(\bm{r}) = \Delta V \sum_{j=0}^{N-1} T(\bm{r} - \bm{s}_j) \: \rho(\bm{s}_j)
\end{equation}
where $S(\bm{r})$ is a $128^2$-element vector of the measured force-gradient signal; 
$N$ is the number of grid points in the spin density map; 
$T(\bm{r} - \bm{s}_j)$ is the point-spread function written in matrix form; 
$\rho(\bm{s}_j)$ is the spin density written as a vector; 
and $\Delta V = \Delta x \: \Delta y \: \Delta z$ is the volume of an individual pixel, with $\Delta x$ and $\Delta y$ from the signal map $S$ and $\Delta z$ the thickness of the grid used to represent the spin-density map $\rho$. 
The point-spread function (psf) $T(\bm{r})$ can be viewed as the response function due to one single spin located at the center of the scanned map, transforming a single spin into a ring of signal. 
As seen in Eq.~\ref{eqn:conv}, both the density $\rho(\bm{s})$ and the psf $T(\bm{r}-\bm{s})$ contain part of the spin information, with the density $\rho(\bm{s})$ carrying the number of spins and their locations while the psf carries the spin's resonance response.
Comparing Eq.~\ref{eqn:conv} and Eq.~\ref{eqn:delta_k_disc}, we obtain the following expression for the 2D psf:
\begin{multline}
\label{eqn:psf}
T(x,y,z - z_0) = \frac{2\pi f_c \Delta t}{\pi^2 x_{\mathrm{0p}}^2} \sum_{q=0}^{N_t-1} \Delta \mu_z (t_q) \\ \times G_x(-x-x_c(t_q),-y,z_0 - z) \: x_c(t_q),
\end{multline}
where $\Delta t = 1/f_c(N_t - 1)$ and the sum implements trapezoid-rule integration over the cantilever oscillation;
$\bm{r}(t_q) = (x + x_c(t_q), y, z)$ is the location of the spherical magnet tip at time $t_q$ during the cantilever oscillation;
 and $z_0$ is a plane of interest in which we generate the psf.
The psf in Eq.~\ref{eqn:psf} is written, for simplicity, for an electron located at the origin.
By varying the value of $z_0$, we obtain 2D psf's for spins located at different vertical distances below the tip.

\begin{figure*}
\includegraphics
	[width=0.7\widthoftwocolumns]
	{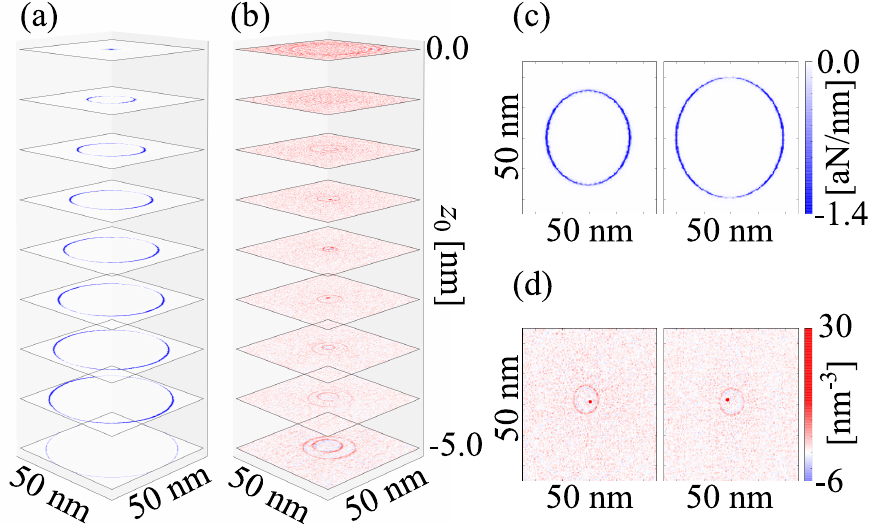}
\caption{(a) A collection of two-dimensional point-spread functions for electron spins located at different distances from the surface of the magnet tip, from 55 \si{\nano\meter} (bottom plane) to 60 \si{\nano\meter} (top plane). This distance range corresponds to $z_0$ planes between -5.0 and 0.0 \si{\nano\meter}. 
Simulation parameters: 
$p_{\mathrm{e}} = 0.32$ and 
$\tau_{\mathrm{p}} = \SI{0.8}{\micro\second}$.
The psf is simulated following the protocol of Fig.~\ref{fig:modulation}b using the steady state result. 
(b) Reconstructed 2D spin density maps at different assumed $z_0$-coordinate using the corresponding psf from (a) and the Fourier deconvolution protocol with Tikhonov regularization. 
(c) The psf's at plane $z_0 = -1.80$ \si{\nano\meter} (left) and $z_0 = -3.05$ \si{\nano\meter} (right) from the tip. 
(d) The reconstructed 2D spin density maps using the corresponding psf's $T$ from (c). 
The Tikhonov regularization parameter was $\lambda = 0.026$ \si{\atto\newton^2 \nano\meter^2}, and the assumed thickness of the spin density map was $\Delta z = 0.08$ \si{\nano\meter}}.
\label{fig:PSF_vs_h}
\end{figure*}

Figure \ref{fig:PSF_vs_h}a shows simulated 2D point-spread functions for a nitroxide spin probe at distances $|z_0 - z|$ ranging from $55$ to $\SI{60}{\nano\meter}$.
Similar to the simulation of the scanned signal, here we kept the microwave irradiation duration to $\tau_{\mathrm{p}} = 0.8$ \si{\micro\second}, and assumed a 2D scanning grid of 128 $\times$ 128 covering an area of 50 \si{\nano\meter} $\times$ 50 \si{\nano\meter}.
The radius of each psf ring in Fig.~\ref{fig:PSF_vs_h}a increases with the distance $|z_0-z|$ from the bottom of the tip, \textit{i.e.}, is different for spins in distinct $z_0$ planes. 
The further the $z_0$ plane is from the tip, the smaller the radius of the psf ring.
Once a $z_0$ plane is further than the deepest point of the resonant slice, where we set the $z =0$ origin, we observe no signal since a spin in that plane cannot intersect with the resonant slice.
The unique shape and size of these psf's can be used to reconstruct the spin density from the measured signal.

\subsection{Fourier deconvolution with Tikhonov regularization}

\begin{figure}
\includegraphics[width=\widthofonecolumn]{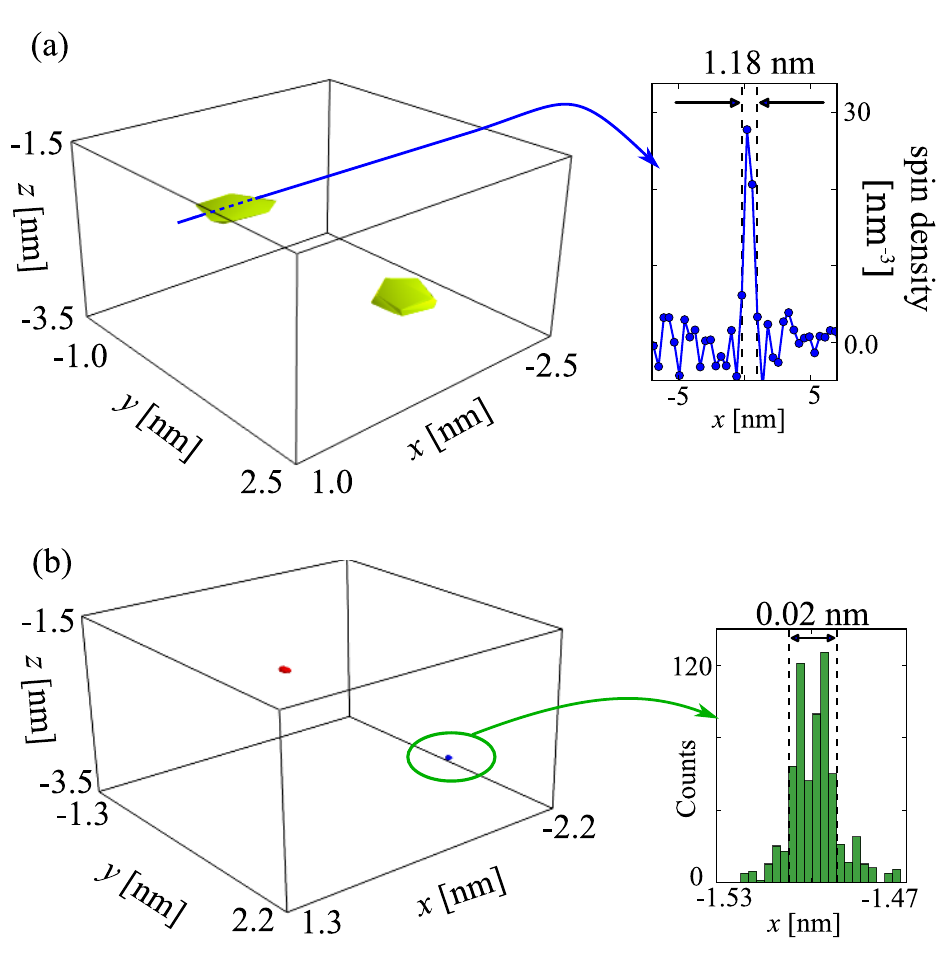}
\caption{Image reconstruction from simulated force-gradient signal using Fourier deconvolution stabilized by Tikhonov regularization followed by reverse Monte Carlo reconstruction. 
(a) Reconstructed three-dimensional image from fast Fourier deconvolution with Tikhonov regularization parameter $\lambda = 0.026 \: \si{\atto\newton^2 \nano\meter^2}$ and an assumed thickness of the spin density map $\Delta z = 0.08$ \si{\nano\meter}. 
The spin density map shows the approximate location of two individual electron spins separated by 21 \si{\angstrom}. 
(inset) Line cut through the reconstructed spin density showing a resolution of $\sim$ 1.18 \si{\nano\meter}. 
(b) The reverse Monte Carlo reconstruction result, sped-up by using the deconvolution result in (a) as an initial guess.
The step size was $\sigma_{\delta\bm{R}} = 0.004$ \si{\nano\meter} and number of iterations was $2\times 10^3$. 
(inset) Posterior distribution for one of the electron spins showing a resolution of $\sim$ 0.2 \si{\angstrom}.}
\label{fig:reconst_Tikh}
\end{figure}

Consider using Eq.~\ref{eqn:conv} to obtain the spin density $\rho$ from the measured signal $S$. For a $N\times N\times N$ discretized sample gird, $\rho$ can be represented as a $N^3$ vector. 
For Eq.~\ref{eqn:conv} to be \textit{prima facia} invertible, the signal $S$ should also be a $N^3$ vector.
In this case the psf $T$ is an $N^3 \times N^3$ circulant matrix. 
Applying the Fourier convolution theorem to Eq.~\ref{eqn:conv}, we can write the Fourier component $\tilde{\hat{S}}_{\ell}$ of the calculated signal $\hat{S}$ as the product of the Fourier components $\tilde{\hat{\rho}}_{\ell}$ and $\tilde{T}_{\ell}$ of the reconstructed spin density $\hat{\rho}$ and the point-spread function $T$, respectively.
\begin{align}
\label{eqn:Fourier_conv}
\tilde{\hat{S}}_{\ell} = \Delta V\, \tilde{T}_{\ell}\, \tilde{\hat{\rho}}_{\ell}
\end{align}

To directly reconstruct a 3D spin density map $\hat{\rho}_{\mathrm{3D}}$ using Eq.~\ref{eqn:Fourier_conv}, we would need a 3D signal map $S(\bm{r})$ with $N^3$ data points.
Here we show that, remarkably, the sparse and localized nature of the electron spin density gives us a way to use Eq.~\ref{eqn:conv} to obtain $\hat{\rho}_{\mathrm{3D}}$ for only $N^2$ measurements of $S(\bm{r})$.
We obtain a 3D approximation to $\hat{\rho}_{\mathrm{3D}}$ (in units of \si{\nano\meter^{-3}}) by aggregating multiple reconstructed 2D spin density arrays $\hat{\rho}_{\mathrm{2D}}$ using an assumed thickness $\Delta z$  based on the distance between the $z_0$ planes: $\hat{\rho}_{\mathrm{3D}} = \hat{\rho}_{\mathrm{2D}} (z_0) /\Delta z$.
By fixing a value of the spin depth $z_0$ in Eq.~\ref{eqn:psf}, the spin density $\hat{\rho}$ in Eq.~\ref{eqn:Fourier_conv} is now an $N^2$ vector $\hat{\rho}_{\mathrm{2D}}$ in units of \si{\nano\meter^{-2}} and can be reconstructed from a 2D signal $S$ using a 2D psf $T$ from Eq.~\ref{eqn:psf}.
As one does in a least-squares fitting problem, 
in this reconstruction method we minimize the non-normalized $\chi^2$ squared difference between the calculated signal $\hat{S}(\bm{r})$ and the measured signal $S(\bm{r})$ in order to obtain the 2D spin density $\hat{\rho}_{\mathrm{2D}}$.
We take $\chi^2$ to be
\begin{equation}
\label{eqn:chi2}
\chi^2(\lambda) = \sum_{i=0}^{N-1} |S_i - \hat{S}_i(\hat{\rho}_{\mathrm{2D}})|^2 + \lambda \sum_{i=0}^{N-1}|\hat{\rho}_i|^2
\end{equation}
with $\hat{\rho}_i = \hat{\rho}_{\mathrm{2D}}(\bm{s}_i)$ the value of the spin density at location $\bm{s}_j$ in the reconstructed image $\hat{\rho}_{\mathrm{2D}}$.
The second term in Eq.~\ref{eqn:chi2} is a regularization factor that we add as a penalty term to keep the spin density small.
Adding this term is valid because we are working with a signal known to arise from only a few spins.
The regularization parameter $\lambda$ has units of $\si{\atto\newton^2 \nano\meter^2}$.

To minimize $\chi^2(\lambda)$, we first apply the Fourier convolution theorem from Eq.~\ref{eqn:Fourier_conv} to $\hat{S}$ in Eq.~\ref{eqn:conv} and substitute the discrete Fourier components of the reconstructed spin density $\hat{\rho}_{\mathrm{2D}}(\bm{s}_i)$ and the signal $S(\bm{r})$, then solve the equation $d\chi^2/d\tilde{\hat{\rho_i}} = 0$ for the spin density in the Fourier domain $\tilde{\hat{\rho}}_{\ell}$
\begin{equation}
\tilde{\hat{\rho}}_{\ell} = \frac{\tilde{T}_{\ell}^* \tilde{S}_{\ell}}{\tilde{T}_{\ell}^* \tilde{T}_{\ell} + \lambda}
\end{equation}
where $\tilde{\hat{\rho}}_{\ell}$, $\tilde{T}_{\ell}$ and $\tilde{S}_{\ell}$ are the discrete Fourier transforms of the estimate spin density $\hat{\rho}_{\mathrm{2D}}(\bm{s})$, the psf $T(\bm{r} - \bm{s})$ and the measured signal $S(\bm{r})$, respectively.
Inverse Fourier transforming $\tilde{\hat{\rho}}_{\ell}$ yields the reconstructed 2D spin density map $\hat{\rho}_{\mathrm{2D}}$. 
Here we used the Tikhonov parameter $\lambda$ as a knob to minimize the error for the reconstructed signal, and chose a value that yields a compromise between obtaining the highest possible resolution for the spin density map and maximizing its signal-to-noise ratio.

We now systematically vary $z_0$ to  obtain a different psf $T$ at each $z_0$ value and reconstruct a 2D spin density map $\hat{\rho}_{\mathrm{2D}}$ at the different $|z_0-z|$ distances from the tip. 
Fig.~\ref{fig:PSF_vs_h}b shows multiple 2D spin density images calculated assuming different values of the depth $z_0$, reconstructed from the measured signal (Fig.~\ref{fig:reconst-Bayes}a) using the 2D psf's at corresponding $z_0$-plane in Fig.~\ref{fig:PSF_vs_h}a and the Tikhonov regularization parameter $\lambda = 0.026$ \si{\atto\newton^2 \nano\meter^2}. 
Noticeably, when the rings of the psf $T$ are of the correct radii, the distance between the tip and the psf plane matches that of the tip-spin separation, and the spin image is well-resolved into a red dot. 
However, when the psf rings are not of the right radii, the reconstructed image only consists of blurred rings (centered, we note, at the correct locations of the two spins). 

Figure \ref{fig:PSF_vs_h}c shows the 2D psf's for specific distances $|z_0-z| = 58.20$ \si{\nano\meter} and $56.95$ \si{\nano\meter} from the end of the tip, and Fig~\ref{fig:PSF_vs_h}d shows the corresponding reconstructed spin density map at those distances, with one electron well-resolved in each spin density map. 
For each psf in Fig.~\ref{fig:PSF_vs_h}c at approximately correct distances, one of the two electrons are well resolved as a dot in Fig.~\ref{fig:PSF_vs_h}d, while the other remains out-of-focus.

Aggregating the 2D reconstructed spin density maps from Fig.~\ref{fig:PSF_vs_h}b, we obtained an approximate 3D spin map, reconstructed from a 2D signal scan. 
Figure \ref{fig:reconst_Tikh}a shows the result of the Fourier deconvolution reconstruction -- a three-dimensional spin density map with resolution of 1.18 \si{\nano\meter}, aggregated from $65$ two-dimensional planes of spin density $\hat{\rho}_{\mathrm{2D}}$ (Fig.~\ref{fig:PSF_vs_h}b) with an assumed thickness of $\Delta z = 0.08$ \si{\nano\meter} between the 2D planes.
The reconstruction using a specific psf was very fast ($\sim 4$ \si{\second}) and the scan of $65$ different values of $|z_0-z|$ took only 5 minutes.
From this 3D spin density map, we extract the electron spins' locations as ($0,0.5,-1.9$) and ($-1.5,1.5,-3$) in units of \si{\nano\meter}, within a few \si{\angstrom} of their true locations of ($0.150, 0.659, -1.814$) and ($-1.505, 1.313, -3.013$) in units of \si{\nano\meter}.
By applying the reverse Monte Carlo reconstruction with the Fourier deconvolution result as the starting point, we now converged with only $2\times 10^3$ iterations, and the reconstruction time was only $\sim \: 13$ minutes, an order of magnitude less than the previous case. 
The stepping distribution of $\delta\bm{R}$ for this sped-up reverse Monte Carlo reconstruction was set to $\sigma_{\delta \bm{R}} = 0.004$ \si{\nano\meter} to allow more detailed sampling of the distribution of the the spins' coordinates.
The acceptance ratio was $\sim 26\%$.
The final distribution of the $x$-coordinate of one electron spin in Fig.~\ref{fig:reconst_Tikh}b shows a resolution of $0.2$ \si{\angstrom}.
The final coordinates of the two electron spins with the sped-up reconstruction are ($0.179 \pm 0.010, 0.666 \pm 0.014, -1.811 \pm 0.002$) and ($-1.525\pm 0.009, 1.322\pm 0.010, -3.013 \pm 0.002$) in units of \si{\nano\meter}. 
The measured distance between the two electron spins from the reconstructed spin map is therefore $2.186 \pm 0.012$ \si{\nano\meter}, which is in quantitative agreement with the known distance of $2.146$ \si{\nano\meter}.
This result is reconstructed from a simulated force-gradient map with an averaging time of $T_{\mathrm{avg}} = 3$ \si{\second} per data point, corresponding to an acquisition time of 13.6 hours for a $128 \:\times \:128$ grid signal map, and a signal-to-noise ratio $\mathrm{SNR} \approx 2$ per point.
Thus, with the Fourier deconvolution result, we were able to bring the starting point for the reverse Monte Carlo reconstruction $\bm{R}_0$ much closer to the spins' actual location, which both sped up the reconstruction process and enhanced the obtained resolution of the spins' coordinates.
The total time required for both Fourier deconvolution and the sped-up reverse Monte Carlo reconstruction was $\sim 18$ minutes, a factor of 10 faster than the reconstruction starting at a random location.

\section{Conclusion}
\label{sec:conclusion}

In conclusion, we have proposed and simulated a mechanical-detection protocol for acquiring magnetic resonance images of individual nitroxide spin labels and have introduced a reverse Monte Carlo method for reconstructing electron coordinates from the expected signal.


The signal-acquisition protocol builds on the idea, demonstrated by Garner \emph{et al.} \cite{Garner2004jun} for nuclear spins and Moore \emph{et al.} \cite{Moore2009dec} for electron spins, of observing spin magnetic resonance as a change in the mechanical oscillation frequency of a magnet-tipped microcantilever.
Detecting this change requires operating at large cantilever amplitude which blurs the signal from individual spins in a scanned-tip imaging experiment.
Here we propose that high spatial-resolution imaging can be retained if the microwave irradiation used to flip sample spins is applied in a short burst delivered in sync with the oscillation of the driven cantilever.
Simulated frequency-shift maps show individual signal rings whose center and radius are directly related to the coordinates of the individual nitroxide spins in the sample.
Using reasonable estimates for the cantilever frequency noise expected near a surface in a magnetic resonance force microscope experiment, we predict that individual electron-spin signals can be detected on the few-seconds timescale and spatially resolved with sub-angstrom resolution on the few hours timescale.
While our simulations were carried out for two spin labels, the protocol should work equally well on scores of spins.


To recover electron-spin coordinates from the expected signal rings we introduced a reverse Monte Carlo algorithm.
The new algorithm has a number of distinct advantages over previously implemented image-reconstruction protocols.
The algorithm's main advantage is that it enables the reconstruction of the three-dimensional distribution of electron coordinates from a two-dimensional signal map, achieving a coordinate uncertainty smaller than the scanning step size and reducing the acquisition time by one to two orders of magnitude.
In contrast with iterative Landweber algorithms \cite{Landweber1951jul,Chao2004may,Degen2009feb}, the new approach is fast, converges, and has a well-defined convergence criterion.
Prior Bayesian Markov-chain Monte Carlo methods introduced by Hero and coworkers assumed a fixed grid of spin density, constrained the total number of spins, but allowed the spin density at each grid point to vary at random \cite{Ting2009jun, Dobigeon2009sep, Park2011jan, Park2012sep, Park2014jan}.
Here we explicitly fix the number of spins and let the spins' coordinates vary stochastically to achieve a maximally probable agreement between observed and calculated signals.


While the performance of our algorithm was demonstrated using a simulated magnetic resonance force microscope signal, the algorithm is applicable to essentially any scanned-tip magnetic resonance experiment operating at the single-spin limit. 
To achieve sub-angstrom precision, the algorithm requires an input consisting of signal rings of a few-nanometer width and only modest signal-to-noise ratio.
The ringed shape of the signal in Fig.~\ref{fig:reconst-Bayes}(a) is primarily determined by the near-spherical nature of the resonant slice.
This spherical shape reflects the geometry of the tip's constant-field surface, which would be the same in any scanned-tip experiment, including those employing nitrogen-vacancy \cite{Grinolds2013feb,Grinolds2014mar} and SQUID detectors.

The work described here was motivated by the exciting possibility of determining the tertiary structure of an individual biomolecular complex by imaging multiple nitroxide spin labels affixed to it \emph{via} uniform labeling \cite{Moore2009dec}. 
We believe that the reduction-of-dimensionality benefit of our algorithm can be realized in a diverse range of single-spin experiments to bring this goal within reach on a practical timescale.

\begin{acknowledgments}

This work was funded by the Army Research Office (Grant No. W911NF-12-1-0221) and Cornell University.

\end{acknowledgments}

\appendix
\section{Spin noise}
\label{sec:supplementary2}

In the protocol of Fig.~\ref{fig:modulation}c we detect the Curie-law magnetization $M_z$.  
Spin fluctuations are a potential source of noise in the Fig.~\ref{fig:modulation}c experiment. 
In this Appendix we assess the size of this noise by assuming that the longitudinal spin fluctuations follow Poisson statistics \cite{Nichol2013sep} and have a correlation time given by $T_1$, the spin-lattice relaxation time.

The signal in Fig.~\ref{fig:modulation}c is given by
\begin{equation}
\label{eqn:Mz-0}
M_z = \frac{1}{N_{\mathrm{a}}}\sum_{i=0}^{N_{\mathrm{a}}-1}\sum_{j=1}^{N_s} \mu_j^{(i)},
\end{equation}
where $N_s$ is the total number of spins in the sample; index $j$ implements a sum over sample spins; $N_{\mathrm{a}}$ is the number of averages in our experiment; and $\mu_j^{(i)}$ denotes the magnetic moment of spin $j$ at the $i^{\mathrm{th}}$ measurement.
The expectation value of the sample magnetization is given by
\begin{align}
\label{eqn:sample-magn-exp}
\expval{M_z} = \frac{1}{N_{\mathrm{a}}}\sum_{i=1}^{N_{\mathrm{a}} -1} \sum_{j=1}^{N_s} \expval{\mu_j} = N_s m_0 p_{\mathrm{e}},
\end{align}
with $\expval{\mu_j} = \mu^{\mathrm{eq}} = p_{\mathrm{e}} \mu_{\mathrm{e}}$ the equilibrium magnetization of the sample spins, $p_{\mathrm{e}}$ the sample polarization, and $m_0 = \mu_{\mathrm{e}}$  the magnitude of the electron spin magnetic moment.
The variance in the measurement of $M_z$ is given by
\begin{equation}
\mathrm{var}[M_z] 
\label{eqn:var_Mz}
= \expval{M_z^2} - \expval{M_z}^2.
\end{equation}
Substituting the expression for $M_z$ from Eq.~\ref{eqn:Mz-0} and its expectation value $\expval{M_z}$ from Eq.~\ref{eqn:sample-magn-exp} into Eq.\ref{eqn:var_Mz}, we can expand the expression for $\mathrm{var}[M_z]$ to
\begin{align}
\mathrm{var}[M_z] &= \expval{\frac{1}{N_{\mathrm{a}}}\sum_{i=0}^{N_{\mathrm{a}}-1}\sum_{j=1}^{N_s} \mu_j^{(i)} \frac{1}{N_{\mathrm{a}}}\sum_{\ell=0}^{N_{\mathrm{a}}-1}\sum_{k=1}^{N_s} \mu_k^{(\ell)}} \nonumber\\
	&\qquad - (N_s m_0 p_{\mathrm{e}})^2 \nonumber \\
\label{eqn:var_Mz_0}
&= \frac{1}{N_{\mathrm{a}}^2} \sum_{i=0}^{N_{\mathrm{a}}-1} \sum_{\ell = 0}^{N_{\mathrm{a}} - 1}\sum_{j=1}^{N_s} \nonumber\\
	&\qquad \bigg[\delta_{j,k} \expval{\mu_j^{(i)}\mu_j^{(\ell)}} + \sum_{k\neq j}^{N_s} \expval{\mu_j^{(i)}}\expval{\mu_k^{(\ell)}} \bigg] \nonumber\\
	&\qquad - (N_s m_0 p_{\mathrm{e}})^2 .
\end{align}

Here we assume that the random fluctuations of one spin  $\mu_j$ are independent of the fluctuations of another spin $\mu_k$, allowing us to separate the expectation value of their product into a product of two expectation values. 
The expectation value $\expval{\mu_j^{(i)}\mu_j^{(\ell)}}$ is given by $m_0^2 (1-p_{\mathrm{e}}^2) e^{-|i-\ell|\tau/T_1} + m_0^2 p_{\mathrm{e}}^2$, with $p_{\mathrm{e}}$ the sample polarization of electron spins and $\tau$ the relaxation time, following Poisson statistics.
As a result, Eq.~\ref{eqn:var_Mz_0} becomes
\begin{align}
\mathrm{var}[M_z]
&= \frac{1}{N_{\mathrm{a}}^2} \sum_{i=0}^{N_{\mathrm{a}}-1} \sum_{\ell = 0}^{N_{\mathrm{a}} - 1} \nonumber\\
	&\qquad \bigg[N_s m_0^2 (1-p_{\mathrm{e}}^2) e^{-|i-\ell|\tau/T_1} + N_s  m_0^2 p_{\mathrm{e}}^2 \nonumber\\
	&\qquad + N_s(N_s -1) m_0^2 p_{\mathrm{e}}^2 \bigg] - (N_s m_0 p_{\mathrm{e}})^2 \nonumber \\
&= \frac{N_s m_0^2}{N_{\mathrm{a}}^2} (1-p_{\mathrm{e}}^2) \sum_{i=0}^{N_{\mathrm{a}}-1} \sum_{\ell = 0}^{N_{\mathrm{a}} - 1} e^{-|i-\ell|\tau/T_1}
\end{align}
The double sum in the above equation may be computed analytically to give
\begin{multline}
\label{eqn:var_Mz_1}
\mathrm{var}[M_z]
= \bigg[N_{\mathrm{a}}(1-e^{-2\tau/T_1})
    - 2e^{-\tau/T_1}(1- e^{-N_{\mathrm{a}} \tau/T_1})\bigg] \\
	\times \frac{N_s m_0^2}{N_{\mathrm{a}}^2}  
	\frac{(1-p_{\mathrm{e}}^2)}{(1-e^{-\tau/T_1})^2}.
\end{multline}
Since $N_{\mathrm{a}} \gg 1$, we drop the second term in the square bracket, yielding
\begin{align}
\mathrm{var}[M_z] 
&=  \frac{N_s m_0^2}{N_{\mathrm{a}}^2} (1-p_{\mathrm{e}}^2) \frac{N_{\mathrm{a}}(1-e^{-2\tau/T_1})}{(1-e^{-\tau/T_1})^2}  \\
\label{eqn:var_Mz_2}
&\approx \frac{N_s m_0^2}{N_{\mathrm{a}}} (1-p_{\mathrm{e}}^2) \frac{2T_1}{\tau}.
\end{align}
In writing Eq.~\ref{eqn:var_Mz_2} we invoke the approximation $e^{-\tau/T_1} \approx 1 - (\tau/T_1)$, valid since the relaxation time in Fig.~\ref{fig:modulation}c is much less than the spin-lattice relaxation time, $\tau/T_1 \ll 1$.
The signal-to-noise ratio for detecting Curie-law magnetization in the limit that the only source of noise is spin fluctuations is obtained by dividing the Curie-law magnetization by the magnetization's standard deviation: 
\begin{equation}
\label{eqn:SNR-ESR-0}
\mathrm{SNR} = \frac{\expval{M_z}}{\sqrt{\mathrm{var}(M_z)}}. 
\end{equation}
Substituting the expression for the expectation value $\expval{M_z}$ from Eq.~\ref{eqn:sample-magn-exp} and the variance $\mathrm{var}[M_z]$ from Eq.~\ref{eqn:var_Mz_2} into Eq.~\ref{eqn:SNR-ESR-0}, we obtain the following  analytic expression for the signal-to-noise ratio
\begin{equation}
\mathrm{SNR} = \frac{N_s m_0 p_{\mathrm{e}}}{\sqrt{\dfrac{N_s m_0^2}{N_{\mathrm{a}}} (1-p_{\mathrm{e}}^2) \dfrac{2T_1}{\tau}}}
\label{eqn:SNR-ESR-imaging}
= \sqrt{\frac{N_{\mathrm{a}}}{N_s}} \frac{p_{\mathrm{e}}}{\sqrt{1-p_{\mathrm{e}}^2}} \sqrt{\frac{\tau}{2T_1}}.
\end{equation}

In the experiment discussed in the main text the number of averages per data point is $N_{\mathrm{a}} = T_{\mathrm{avg}}/(n T_c) \approx 2\times 10^4$.
This number is computed assuming an averaging time of $T_{\mathrm{avg}} = 3$ \si{\second} per point, a typical cantilever oscillation period of $T_c = 154$ \si{\micro\second}, and $n = 1$. 
When $n=1$, the relaxation time is $\tau = T_c = 154$ \si{\micro\second}.
The electron spin-lattice relaxation time is $T_1 = 1.3$ \si{\milli\second} and the 
number of electron spins in the sample is $N_s = 2$.
At a temperature $T_0 = 2.1$ \si{\kelvin} and field $B_0 = 1.4$ \si{\tesla}, the thermal polarization of the electron spins is $p_{\mathrm{e}} = p^{\mathrm{therm}} = 0.42$. 
Substituting these experimental parameters into Eq.~\ref{eqn:SNR-ESR-imaging} we obtain a signal-to-noise ratio of $\mathrm{SNR} \approx 11$.
This SNR is much higher than the $\mathrm{SNR} \approx 2$ seen in the experiment considered in the main text in which thermal detector fluctuations were the main source of noise.
We conclude that the experiment of Fig.~\ref{fig:modulation}c is operating in the detector noise limit as was assumed.

\section{Numerical Bloch simulation}
\label{sec:supplementary4}

\begin{figure}
\includegraphics[width=\widthofonecolumn]{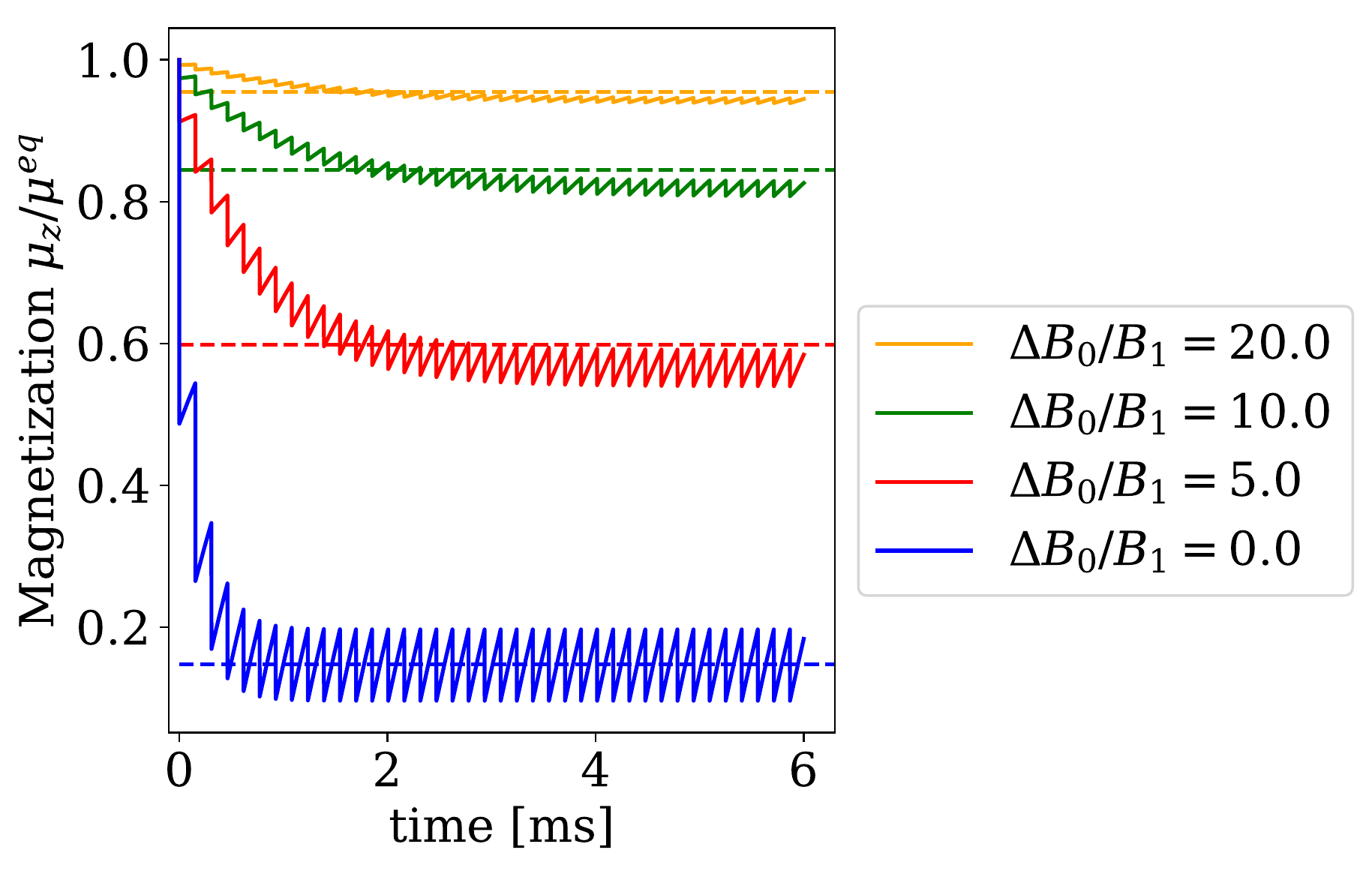}
\caption[Cyclic saturation with intermittent microwave irradiation]{Numerically simulated longitudinal magnetization $\mu_z$ \emph{vs.} time in the Fig.~\ref{fig:modulation}c experiment.
The magnetization is computed at various resonance offsets (solid lines).
The steady-state longitudinal magnetization predicted by Eq.~\ref{eqn:response-lineshape} is plotted for comparison (dashed lines).
}
\label{fig:steady-state-Bloch}
\end{figure}

The spin magnetization $\mu_z$ in the cyclic saturation experiment of Fig.~\ref{fig:modulation}c was simulated by numerically integrating the Bloch equations using the Python package \texttt{scipy.integrate.odeint}.
The electron-spin resonance response function shown in Fig.~\ref{fig:Bloch_numerical}, described by Eqs.~\ref{eqn:response-lineshape} and \ref{eqn:halfwidth}, was obtained \emph{via} numerical simulation as follows.
At each resonance offset the evolution of the vector magnetization was computed for 100 cycles of evolution where each cycle consisted of a microwave burst (duration $\tau_{\mathrm{p}} = \SI{0.8}{\micro\second}$, represented by $101$ time points) followed by a relaxation delay  (duration $\tau = \SI{154}{\micro\second}$, represented by $192$ time points).
The total number of time points in each simulation was $N_{\mathrm{pts}} = 29,\!300$ and each simulation took 3 seconds to execute.
To simplify the simulation, the $x$ and $y$ components of the magnetization were set to zero at the end of the relaxation delay; this is a valid approximation because $\tau \gg T_2$. 
Representative magnetization \emph{vs.} time traces are shown in Fig.~\ref{fig:steady-state-Bloch} for four different resonance offsets ranging from $\Delta B_0 = 0$ to $\SI{200}{\micro\tesla}$.
We can see that $\mu_z$ reaches a well-defined steady-state value when subjected to intermittent irradiation that is well described by Eqs.~\ref{eqn:response-lineshape} and \ref{eqn:halfwidth}.


\label{TheEnd}
\end{document}